\documentclass{article}

\usepackage{amsmath}

\usepackage[numbers]{natbib}
\usepackage{arxiv}

\usepackage{mathtools}
\usepackage{amsfonts}
\usepackage{amssymb}
\usepackage{amsthm}
\usepackage{mathrsfs}
\usepackage{enumitem}
\usepackage{xcolor}
\usepackage{xpatch}
\usepackage{booktabs}
\usepackage{xr}

\usepackage[plain,noend]{algorithm2e}

\makeatletter
\renewcommand{\algocf@captiontext}[2]{#1\algocf@typo. \AlCapFnt{}#2} 
\def\@algocf@capt@plain{top}
\renewcommand{\algocf@makecaption}[2]{%
  \addtolength{\hsize}{\algomargin}%
  \sbox\@tempboxa{\algocf@captiontext{#1}{#2}}%
  \ifdim\wd\@tempboxa >\hsize
    \hskip .5\algomargin%
    \parbox[t]{\hsize}{\algocf@captiontext{#1}{#2}}
  \else%
    \global\@minipagefalse%
    \hbox to\hsize{\box\@tempboxa}
  \fi%
  \addtolength{\hsize}{-\algomargin}%
}
\makeatother

\def\T{{ \mathrm{\scriptscriptstyle T} }}

\let\longto\longrightarrow
\DeclareMathOperator{\cov}{cov}
\DeclareMathOperator{\Tr}{tr}
\DeclareMathOperator{\var}{var}

\DeclareMathOperator{\G}{\mathcal{G}}

\DeclareMathOperator{\A}{\mathcal{A}}
\DeclareMathOperator{\B}{\mathcal{B}}
\def\T{{ \mathrm{\scriptscriptstyle T} }}

\newtheorem{lemma}{Lemma}
\newtheorem{theorem}{Theorem}

\title{Generalizations to Corrections for the Effects of Measurement Error in Approximately Consistent Methodologies}

\date{}

\author{
	D. Spicker\\
	Statistics and Actuarial Science\\ 
	University of Waterloo\\ 
	Waterloo, Ontario, N2L 3G1\\
	\texttt{dylan.spicker@uwaterloo.ca} \\
	\And
	M. P. Wallace \\
	Statistics and Actuarial Science\\ 
	University of Waterloo\\ 
	Waterloo, Ontario, N2L 3G1 \\
    
	\And
	G. Y. Yi \\
	Department of Statistical and Actuarial Science\\ 
    Department of Computer Science\\
    Western University, \\ 
    London, Ontario, N6A 5B7
}

\begin{document}
\markboth{D. Spicker et~al.}{Generalizations of Approximately Consistent Measurement Error Models}
\maketitle

\begin{abstract}
    Measurement error is a pervasive issue which renders the results of an analysis unreliable. The measurement error literature contains numerous correction techniques, which can be broadly divided into those which aim to produce exactly consistent estimators, and those which are only approximately consistent. While consistency is a desirable property, it is typically attained only under specific model assumptions. Two techniques, regression calibration and simulation extrapolation, are used frequently in a wide variety of parametric and semiparametric settings. However, in many settings these methods are only approximately consistent. We generalize these corrections, relaxing assumptions placed on replicate measurements. Under regularity conditions, the estimators are shown to be asymptotically normal, with a sandwich estimator for the asymptotic variance. Through simulation, we demonstrate the improved performance of the modified estimators, over the standard techniques, when these assumptions are violated. We motivate these corrections using the Framingham Heart Study, and apply the generalized techniques to an analysis of these data.
\end{abstract}


\section{Introduction}
\label{sec::introduction}
Variables measured with error often pose a significant concern for valid inference. Any statistical analysis which relies on error-prone variables will be impacted in a manner that depends on the structure and size of the measurement error, and on the analysis technique that is used. A large number of techniques have been proposed in order to correct for the effects of measurement error. These correction techniques typically rely on auxiliary information, often in the form of replicate, or repeated, measurements of the variable of interest. We consider two correction techniques, regression calibration \citep{Carroll1990,Gleser1990} and simulation extrapolation (SIMEX, \cite{SIMEX}), which are often applied when replicate measurements are available. We investigate these techniques in the setting where the repeated measurements that are available are not truly replicate measurements. In particular, we consider the setting where the available measurements are not assumed to be identically distributed, and demonstrate the ease with which these correction techniques can be generalized for non-identically distributed data.

There is a large body of research focused on correcting for the effects of measurement error, in specific models, with an emphasis on producing consistent estimators. Consistency is a desirable property which often comes at the cost of specific model assumptions and complex implementation, thus limiting the scope of application. Violations of these assumptions may severely undermine the utility and validity of these methods \citep{CarrollBook,SIM_tutorial,YiBook}. On the contrary, \emph{approximately consistent} methods trade rigorous mathematical guarantees for more general utility and ease of implementation. Our focus on regression calibration and SIMEX is due to the comparative ease of implementation and general utility of these methods, which has led to their widespread use in applied settings \citep{stratos}.

A recent survey examined studies related to dietary patterns, physical activity, and air pollution, areas which are known to be impacted by measurement error. The results of this survey conclude that ``[...] while researchers were generally aware that measurement error affected their studies, very few adjusted their analysis for the error. [...] Regression calibration was the most widely used method of adjustment.'' \citep{stratos}. These survey results suggest that it is important to develop theoretically justified methodologies, which accommodate the wide variety of data structures that may arise in applied scenarios. Our focus is on expanding the utility of regression calibration and SIMEX, without changing the underlying mechanics of implementation, allowing for analysts familiar with the tools to incorporate these results.

As a motivating example, we consider the Framingham Heart Study (FHS, \cite{Framingham}). The Framingham Heart Study is a large cohort study investigating coronary heart disease (CHD). Following previous analyses in the measurement error literature \citep{CarrollBook}, our interest concerns the impact of long-term average systolic blood pressure (SBP) on the development of coronary heart disease, while controlling for other risk factors. Systolic blood pressure is considered error-prone since it is a single time point measurement which may differ substantially from the long-term average blood pressure. The study reports multiple separate measurements of systolic blood pressure, from each clinical visit. In their analysis, \citeauthor{CarrollBook} treat these measurements as replicates of the same underlying value. That is, the systolic blood pressure measurements are considered independent and identically distributed (iid) realizations of the same variable. They note that these sets of measurements significantly deviate from one another, however, these differences are not material enough to meaningfully change the estimated parameter values. We will consider a similar analysis of a different subset of the FHS data, where once again the assumption that the separate measurements are replicates is violated, but where these violations are not readily accommodated by the existing techniques.

\section{Methodology}\label{sec::methods}
We begin by introducing the typical modeling assumptions made when using regression calibration and SIMEX, before describing the implementation of these methods. We use $i$ to index over observations, and for the $i$-th observation define $X_i$ to be the true value for our variate of interest, $X_i^*$ to be its observed version or surrogate value, $Y_i$ to be a numeric outcome, and $Z_i$ to represent all other covariates which are measured without error.

\subsection{Measurement Error Background}
Measurement error is often assumed to follow an additive model. The \emph{classical additive model} posits that we observe $X_i^* = X_i + U_i$, where we assume $E[U_i] = 0$, $\var(U_i) = \sigma_U^2$ is constant with respect to $X_i$, $U_i$ is independent of $X_i$ (denoted $U_i\perp X_i$), and $U_i\perp(Y_i, Z_i)$. The surrogate $X_i^*$ is therefore an unbiased measurement of $X_i$, in the sense that $E[X_i^*|X_i] = X_i$. Often, $U_i$ is assumed to follow a normal distribution. 

We may also consider \emph{multiplicative models} which typically assume the form $X_i^* = X_iV_i$, with $E[V_i] = 1$, constant variance given by $\var(V_i) = \sigma_{V}^2$, $V_i\perp X_i$, and $V_i\perp(Y_i,Z_i)$. Thus, $X_i^*$ is unbiased for $X_i$. Much of the methodology developed in the literature assumes an additive structure, and it is often advised to transform a measurement that has multiplicative error onto a scale where the error becomes additive \citep{CarrollBook,Multiplicative_Transformations}. 

Our notation implies that the random variables are scalar valued. Vector valued random variables can be accommodated by allowing each component to follow analogous expressions. For some results, it will be convenient to unify this framework notationally. We can write $X_i^* = X_i + \delta U_i$ or $X_i^* = X_i(1 + \delta U_i)$, where $U_i$ is a mean-zero, unit variance random variable, independent of $X_i$, $Y_i$ and $Z_i$, and $\delta$ is constant.

In order to correct for the effects of measurement error, its size must be assessed, and methods require additional information to do so. In addition to the assumed measurement error model, applicable methods are often dictated by what auxiliary data are available. These may come in the form of \emph{validation} data, where for some individuals both $X_i$ and $X_i^*$ are available, the aforementioned replicate data, where repeated iid measurements of $X_i^*$ are taken, or \emph{instrumental} data. Instrumental data involve the measurement of $X_i^*$ and an additional variable $T_i$, called the instrumental variable, which is related to both the outcome, $Y_i$, and the surrogate, $X_i^*$, only through the true value, $X_i$. Replicate data are a specific form of instrumental data.

Validation data are typically preferred, but are often unavailable or impossible to collect. In the FHS, for example, long-term average SBP is desired, which cannot be measured in practice. There is literature which uses instruments for correction of measurement error effects, some of which extends the idea of regression calibration \citep{InstrumentalBackground,CarrollBook}. However, the conditions required for a variable to be viewed as an instrumental variable are often difficult to verify in practice, and falsely assuming that a factor is an instrument can lead to large biases, even when the bias introduced from the measurement error is small to begin \citep{CarrollBook}. Consequently, most of the literature has focused on the use of replicate measurements. When using replicate data, we assume that we have a sample in which we observe $\{Y_i, Z_i, X_{i1}^*,\hdots,X_{i\kappa_i}^*\}$, where $\{X_{i1}^*,\hdots,X_{i\kappa_i}^*\}$ are $\kappa_i$ iid replicates from the same error model. 

\subsection{Regression Calibration (Best Linear Unbiased Prediction Technique)}\label{subsec::rc}
Regression calibration \citep{Carroll1990,Gleser1990} models the relationship between $Y_i$ and $\{X_i^*, Z_i\}$ based on an assumed model between $Y_i$ and $\{X_i,Z_i\}$ and a model between $X_i$ and $\{X_i^*,Z_i\}$. While there exist many specific implementations of regression calibration, we consider the \emph{best linear unbiased prediction} (BLUP) technique \citep{CarrollBook}. 

The BLUP technique assumes that there is a standard model, informed by the underlying modeling subject matter, which relates $Y_i$ to $\{X_i,Z_i\}$, and that $E[X_i|X_i^*,Z_i]$ is well represented by a linear structure. Then, we replace $X_i$ in the standard model with $\widehat{X}_i \triangleq \widehat{E}[X_i|X_i^*,Z_i]$, representing an estimate of $E[X_i|X_i^*,Z_i]$. If we assume that replicate measurements are available, we define the BLUP of $X_i$, as the linear quantity which minimizes the mean-squared error (MSE), when viewed as a predictor of $X_i$. That is, we find $\widehat{X}_i = \mu + \beta X_i^* + \gamma Z_i$ such that $E[(X_i - \widehat{X}_i)^2]$ is minimized, with respect to $\mu$, $\beta$, and $\gamma$. Taking $\mu_A$ and $\Sigma_{AB}$ represent the mean of $A$ and the covariance between $A$ and $B$, we get \begin{equation}
    \widehat{X}_i = \mu_X + \begin{bmatrix}\Sigma_{XX^*} & \Sigma_{XZ}\end{bmatrix}\begin{bmatrix}
        \Sigma_{X^*X^*} & \Sigma_{X^*Z} \\
        \Sigma_{ZX^*} & \Sigma_{ZZ}
    \end{bmatrix}^{-1}\begin{bmatrix}
        X_i^* - \mu_{X^*} \\
        Z_i - \mu_Z
    \end{bmatrix}.\label{eq::standard_blup}
\end{equation}

Under the assumption that replicate data are available, we usually take $X^*_i = \kappa_i^{-1}\sum_{j=1}^{\kappa_i}X_{ij}^*$. All quantities in equation (\ref{eq::standard_blup}) can be estimated using the replicate measurements. In the event that $X_i$ and $U_i$ are jointly normally distributed, this method consistently estimates the conditional mean $E[X_i|X_i^*,Z_i]$. If the joint distribution of $X_i$ and $U_i$ is non-normal, but $\var(U_i)$ is sufficiently small, it has been shown that this technique can provide an acceptable prediction of the conditional mean of $X_i$ given $X_i^*$ and $Z$ \citep{Carroll1990}.

Once we have values for $\widehat{X}_i$, we fit the model explaining $Y_i$ in terms of $\{X_i,Z_i\}$, using $\{\widehat{X}_i,Z_i\}$ instead. To formalize notation we will assume that our interest is in a vector of parameters, $\Theta_0$, which are consistently estimated as the solution to an estimating equation, say, \[0 = U_n(Y,X,Z;\widehat{\Theta}) = \sum_{i=1}^n \Psi(Y_i,X_i,Z_i;\widehat{\Theta})\], where $E[\Psi(Y_i,X_i,Z_i;\Theta)] = 0$ when $\Theta = \Theta_0$. We also call $\widehat{\Theta}$ an \emph{M-estimator}. While the M-estimator approach is not strictly necessary, it should be sufficiently general to cover most estimators of interest. The regression calibration estimator, $\widehat{\Theta}_\text{RC}$, is then the solution to $0 = U_n(Y,\widehat{X},Z; \widehat{\Theta}_\text{RC})$.

If the model for $Y_i$ in terms of $\{X_i,Z_i\}$ is linear, and the BLUP consistently estimates the conditional mean function, $E[X_i|X_i^*, Z_i]$, then the resultant estimator $\widehat{\Theta}_\text{RC}$ will be consistent, provided mild regularity conditions. If the model for $Y_i$ is non-linear, or the mean function is not consistently estimated, we cannot generally guarantee consistency. In certain non-linear models, however, specific claims can be made. If we fit a log-linear model, with the correct mean function, then all slope parameters will be consistently estimated. If we fit a logistic regression model then, generally, bias will be reduced and, if the key interest is in the estimated probabilities rather than the parameters, the approximation will often be quite accurate \citep{CarrollBook}. Thus the regression calibration estimators are generally regarded as approximately consistent \citep{CarrollBook}. Under regularity conditions, after certain transformations, they are asymptotically normal \citep{Carroll1990}.

While the BLUP technique typically assumes that the replicates are iid, regression calibration can still be applied when this is not the case. The correction will work as presented when there is complete replication; that is when $\kappa_i = k$ for all $i$. In order to estimate the parameters required for the BLUP correction, it is necessary to estimate the variance of the measurement error term, denoted $\Sigma_{UU}$. When the replicates are identically distributed, there is a single variance term which is estimable through this method. When the replicates are not identically distributed, there may be $\kappa_i$ separate variances. The standard regression calibration procedure, however, produces an estimator $\widehat{\Sigma}_{UU}$ which is consistent for $\frac{1}{\kappa_i}\sum_{j=1}^{\kappa_i} \Sigma_{U_jU_j}$. Under the assumption of complete replication, this expression will serve as the error variance for $\frac{1}{\kappa_i}\sum_{j=1}^{\kappa_i} X_{ij}^*$, allowing the correction to proceed as outlined.

However, it is often the case that different replicate measurements are observed for each $i$. When this is true, this standard procedure will fail to consistently estimate the error variances, as there is not one error variance to estimate. The evident solution is to, if possible, estimate separate error variances for each repeated measurement. If they are truly identically distributed, then the average of these estimates will serve as an estimator for the single error variance. Otherwise, each variance term can be used, and combined, as is necessary for each $i$ depending on which replicates are observed. In this sense, we can view the standard regression calibration, where we assume that the replicate measurements are identically distributed, as a special case of the more general procedure. One reason that the analysis of the FHS study presented by \citeauthor{CarrollBook} is not impacted by the non-iid nature of the replicates is that the subset used in their analysis has complete replication. The estimated variance is consistent for the average variance, and the correction can proceed. The sample of the FHS that we use has incomplete replication.

\subsection{Simulation Extrapolation}
As its name suggests, the simulation extrapolation \citep{SIMEX} method is divided into two steps: simulation and extrapolation. In the simulation step, data subject to larger measurement errors are simulated, so that the analyst can see the impact of this error on the bias of the estimated parameter of interest. Then, in the extrapolation step, this relationship is extrapolated back to the case where no measurement error is present. We will continue to assume the parameters of interest, $\Theta_0$, are consistently estimated in the error-free setting using an M-estimator. Our presentation of SIMEX will assume that $\Theta_0$ is a scalar, but the method applies in higher dimensions. Assume that $\var(U) = \delta$ is known. For some positive constant $\lambda$, we construct the quantity $X_{bi}^*(\lambda) = X_i^* + \left(\lambda\delta\right)^{1/2}\epsilon_{bi}$, where the $\epsilon_{bi}$ are generated by the analyst to be iid $N(0,1)$ (pseudo-)random variables, independent of the $\{Y_i, X_i^*, X_i, Z_i\}$, called \emph{pseudo-errors}. 

For $\lambda \geq 0$, the variance of $X_{bi}^*(\lambda)$ conditional on $X_i$ is given by $(1 + \lambda)\delta$. If $\delta=0$, then our measured values would not deviate from the truth, and taking $\lambda = -1$ reflects this situation. For any $\lambda \geq 0$, we can generate $\widehat{\Theta}_b(\lambda)$ as the solution to $0 = U_n(Y,X_{b}^*(\lambda),Z; \widehat{\Theta}_b(\lambda))$. Averaging the results over $b=1,\hdots,B$ independent simulations produces an arbitrarily precise (by taking $B$ large) estimator for the quantity $\Theta(\lambda) = E\left[\widehat{\Theta}_b(\lambda)\right]$.

Repeating this process for many values of $\lambda$, say $0 \leq \lambda_1 < \lambda_2 < \cdots < \lambda_R$, generates a sequence of estimators for $\{\Theta(\lambda_1), \hdots, \Theta(\lambda_R)\}$. We can model these values as a function of $\lambda$, $\G(\lambda)$ if we specify a particular parametric form for $\G(\lambda)$. Using least squares estimation we can estimate $\G(\lambda)$ as $\widehat{\G}(\lambda)$. Extrapolation occurs by taking $\widehat{\Theta}_\text{SIMEX} = \widehat{\G}(-1)$, representing the error-free setting. Generally, one of three extrapolants are used: a linear extrapolant, $\G(\lambda) = a_0 + b_0\lambda$, a quadratic extrapolant, $\G(\lambda) = a_0 + b_0\lambda + c_0\lambda^2$, or a nonlinear extrapolant, $\G(\lambda) = a_0 + \frac{b_0}{c_0 + \lambda}$, where $a_0$, $b_0$, and $c_0$ are regression coefficients \citep{SIMEX}.

If the errors are normally distributed, then for a correctly specified, sufficiently smooth $\G$ \citep{SIMEX_jackknife}, $\widehat{\Theta}_\text{SIMEX}$ consistently estimates $\Theta_0$. More broadly, SIMEX can be seen as an approximately consistent estimation technique. It consistently estimates $\lim_{\lambda\to-1} E[\widehat{\Theta}_b(\lambda)]$, which itself approximates the true value $\Theta_0$, assuming correct specification of $\G$. The quality of this approximation will dictate the quality of the estimator. 

The preceding description of SIMEX has assumed knowledge of $\delta$. While this may be available, it is more likely that it must be estimated through auxiliary data. When replicate data are available, and if error variances are assumed to be homogeneous, then it is also possible to use the replicate measurements to estimate the error variance, and use this estimated quantity in place of $\delta$ above. Under regularity conditions, the resultant estimators are asymptotically normal \citep{SIMEX_asym}. When replicates are availabule, but error variances are heterogeneous, a modified version of SIMEX, known as the empirical SIMEX, can be used \citep{SIMEX_empirical}. We focus on the standard SIMEX.

Just as with regression calibration, SIMEX can accommodate non-identically distributed replicates, assuming complete replication. SIMEX will consistently correct for errors, assuming the correct extrapolant and normality of errors, so long as the variance of $X_{bi}^*(\lambda)$ tends to $0$ as $\lambda$ tends to $-1$. This means that, as long as we correctly estimate the error variance, the correction will work, and as a result, the discussion regarding $\widehat{\Sigma}_{UU}$, and the corresponding limitations, from the section~\ref{subsec::rc} holds verbatim.

\section{Generalizations of the Methods}\label{sec::generalized_methods}
As discussed, both methods presented can accommodate non-identically distributed replicate measurements, under certain assumptions. However, treating these techniques as special cases in a slightly broader framework, we can maintain the appeal of both regression calibration and simulation extrapolation, while accommodating this slightly more general error structure. To do so, we present a generalized measurement error model. First we introduce the model and discuss identifiability concerns. Then we illustrate the implementation of, and justification for, regression calibration and SIMEX using this model.

\subsection{Data Structure and Identification}
We will assume that we have observations $X_{ij}^*$, $j=1,\hdots,\kappa_i$, where for each observation we have either an additive structure $X_{ij}^* = \eta_{0j} + \eta_{1j}X_i + U_{ij} = \eta_{0j} + \eta_{1j}X_i + \delta_jU_{ij}$ or a multiplicative structure $X_{ij}^* = \eta_{0j} + \eta_{1j}X_iV_{ij} = \eta_{0j} + \eta_{1j}X_i(1 + \delta_jU_{ij})$, where $U_{ij}$, $V_{ij}$, and $\delta$ are as before. All error processes are assumed to be independent of each other and of $\{Y_i,X_i,Z_i\}$. The multiplicative error case can be made multivariate using Hadamard products, denoted $\circ$. We will refer to this data structure as the \emph{generalized measurement error model}. 

In this setup, taking $I(\cdot)$ to be the indicator function, we have: $E\left[X_{ij}^*\right] = \eta_{0j} + \eta_{1j}\circ E[X_i]$; $\cov\left(X_{ij}^*, X_{il}^*\right) = \eta_{1j}^{(d)}\Sigma_{XX}\eta_{1l}^{(d)} + I(j=l)M_j$; $\cov\left(X_i, X_{ij}^*\right) = \Sigma_{XX}\eta_{1j}^{(d)}$; and $\cov\left(Z_i, X_{ij}^*\right) = \Sigma_{ZX}\eta_{1j}^{(d)}$, where $j=1,\hdots,\kappa_i$, indexes proxy measurements and $i=1,\hdots,n$ indexes the individuals. $\eta_{1j}^{(d)}$ represents the diagonal matrix with the elements of $\eta_{1j}$ along its diagonal. Further, $E\left[\left.X_{ij}^*\right|X_i, Z_i\right] = \eta_{0j} + \eta_{1j}\circ X_i$, and $\var\left(\left.X_{ij}^*\right|X_i\right) \triangleq M_j(X_i)$, where, $M_j$ and $M_j(X_i)$ are matrices that capture the variance of the assumed error model, taking the form of $M_j = M_j(X_i) = \Sigma_{U_{j}U_{j}}$ if an additive structure is assumed, and $M_j = \eta_{1j}^{(d)}\left(E[X_iX_i']\circ\Sigma_{V_jV_j}\right)\eta_{1j}^{(d)}$ or $M_j(X) = \eta_{1j}^{(d)}\left(XX'\circ\Sigma_{V_jV_j}\right)\eta_{1j}^{(d)}$ otherwise.

One consideration which is important concerns the structure of $M_j$ under the multiplicative structure In particular, the matrix is not guaranteed to be positive semidefinite, for multivariate $X_j^*$. This is not a concern in the estimation procedure for $M_j$, but it will impact the ways in which it can be used. In particular, in order to think of $M_j$ as a variance term, we must either limit our discussion to additive measurement error models, or work with univariate variables.

\subsection{Parameter Identification}\label{subsec::parameter_identification}
We must impose restrictions on some model parameters in order to render the model identifiable. We will assume, for some set of $j$, that (1) $\eta_{0j} = 0$, (2) $\eta_{1j} = 1$, or (3) both $\eta_{0j} = 0$ and $\eta_{1j} = 1$. These assumptions also capture the case where, for instance, $\eta_{0j} = c$ for any constant $c$. If $c$ is non-zero, we can work with $X_j^* - c$, leaving us with a measurement satisfying assumption (1). When $\eta_{0j} = 0$ and $\eta_{1j} = 1$ for all $j$, this model reduces to that of having $\kappa_i$ unbiased measurements of $X$, from possibly different distributions. Other assumptions will also suffice.

We define $J_0$, $J_1$, and $J_{01 }= J_0\cap J_1$ to be the index sets for the proxies corresponding to assumptions (1), (2), and (3) respectively. We assume that $|J_0| \geq 1$ and $|J_1| > 1$, which is not necessary, but will suffice for the identification of the parameters. We denote the $k^2$ parameters given by $E[X_j^*]$ and $\cov(X_j^*, X_l^*)$ as $\mu_j$ and $\Sigma_{X_j^*X_l^*}$, respectively.

Assuming that $\kappa_i = k$ for all $i$, we take $\widehat{\mu}_X = |J_{0}|^{-1}(\sum_{j \in J_{01}} \mu_j + \sum_{j\in J_0\setminus J_{01}} \widehat{\eta}_{1j}^{(d)^{-1}}\mu_j)$; $\widehat{\eta}_{0j} = \mu_j - \widehat{\eta}_{1j}^{(d)}\widehat{\mu}_X$; $\widehat{M}_j = \Sigma_{X_j^*X_j^*} - \widehat{\eta}_{1j}^{(d)}\widehat{\Sigma}_{XX}\widehat{\eta}_{1j}^{(d)}$; and $\widehat{\Sigma}_{X} = k^{-1}\sum_{j=1}^k \widehat{\eta}_{1j}^{(d)^{-1}}\widehat{\Sigma}_{XX_j^*}$. If $Z_i$ is not observable, then we can take $\widehat{\Sigma}_{XX_j^*} = |J_1\setminus\{j\}|^{-1}\sum_{l \in J_1\setminus\{j\}} \Sigma_{X_l^*X_j^*}$ and $\widehat{\eta}_{1j}^{(d)} = (K-1)^{-1}\sum_{l\neq j; l=1}^K \Sigma_{X_j^*X_l^*}\widehat{\Sigma}_{XX_l^*}^{-1}$. If $Z_i$ is observable then we take $\widehat{\Sigma}_{ZX} = |J_1|^{-1}\sum_{j\in J_1} \Sigma_{ZX_j^*}$; $\widehat{\eta}_{1j}^{(d)} = \{(\widehat{\Sigma}_{XZ}\widehat{\Sigma}_{ZX})^{-1}\Sigma_{X_j^*Z}\Sigma_{ZX_j^*}\}^{1/2}$; and $\widehat{\Sigma}_{XX_j^*} = (k-1)^{-1}\sum_{l\neq j; l=1}^K \widehat{\eta}_{1l}^{(d)^{-1}}\Sigma_{X_l^*X_j^*}$. With $Z_i$ observable, $|J_1| = 1$ is permissable. We collectively refer to these estimators as the correction parameter estimators. The parameters needed to compute these correction parameter estimators will not typically be known, however, they can be consistently estimated with the observed data.

\begin{lemma}
\label{lemma::m_estimator_parameters}
    The correction parameter estimators can be expressed as the solution to unbiased estimating equations subject to standard asymptotic theory. We can write that $\widehat{\xi} = (\widehat{\mu}_j, \widehat{\Sigma}_{X_j^*X_l^*}, \widehat{\mu}_X, \widehat{\Sigma}_{XX}, \widehat{\eta}_{0j}, \widehat{\eta}_{1j}, \widehat{M}_{j}, \widehat{\Sigma}_{XX_j^*})$ is given by the solution to $0 = n^{-1}\sum_{i=1}^n g(X_i^*, Z_i, \widehat{\xi})$, and for the true value $\xi$, we have $0 = E\{g(X^*,Z,\xi)\}$. The form of $g$ is given in the supplementary material as equation (S5). \textbf{Proof:} See the supplementary material.
\end{lemma}

By altering the form of $g$, this result applies if $Z$ is available. Since we have assumed $\kappa_i = k$ for all $i$, these estimators, as presented, are not applicable for incomplete replication. The estimators can be viewed as a restated version of the standard regression calibration parameter estimators, within a slightly more general framework, where we have explicitly communicated the fact that there is no assumption of identically distributed measurements. These estimators, however, have been constructed in such a way so as to accommodate incomplete replication. If we assume that the set of replicates available for each individual are independent of the measured variables, we can continue in the modified framework. Under this assumption of \emph{ignorable missingness}, the $j$-th proxy's parameters are consistently estimable using only the observations which have the $j$-th proxy available. The function that the M-estimators are based on, $g$, can be modified to include the observation indicators of $X_{ij}^*$. 

Standard asymptotic results demonstrate that $n^{1/2}(\widehat{\xi} - \xi)$ converges in distribution to $N(\mathbf{0}, \A^{-1}(\xi)\B(\xi)\A^{-1}(\xi)^\T)$, as $n\to\infty$. Here $\A(\xi) = E\{\partial/\partial\xi^\T g(X^*,Z,\xi)\}$ and $\B(\xi) = E\{g(X^*,Z,\xi)g(X^*,Z,\xi)^\T\}$, and both are estimable consistently from the data. As outlined in Section~\ref{sec::methods}, we focus on estimation techniques which can be framed as M-estimators. Using Lemma~\ref{lemma::m_estimator_parameters} we can derive the asymptotic distribution for estimators derived from any correction methods that use the contents of $\xi$ and M-estimation. 

\begin{lemma}
\label{lemma::general_asymptotic_distribution}
    Assume that $\widehat{\xi}$ solves the empirical estimating equation from Lemma \ref{lemma::m_estimator_parameters}, denoted $g_n(\widehat{\xi}) = 0$, and that $\widehat{\Theta}$ is a solution to the empirical estimating equation $U_n(\widehat{\Theta},\widehat{\xi}) = 0$, where $\widehat{\xi}$ and $\widehat{\Theta}$ are estimating $\xi$ and $\Theta$, respectively. Then we have that $n^{1/2}(\widehat{\Theta} - \Theta)$ converges in distribution to $N(0, \Sigma_{(1)})$, as $n\to\infty$, where $\Sigma_{(1)} = Q\A^{-1}(\Theta,\xi)\B(\Theta,\xi)A^{-1}(\Theta,\xi)^\T Q^\T$, for $Q = \begin{bmatrix}I_{p\times p} & 0_{p\times q}\end{bmatrix}$, $\A(\Theta,\xi)$ is upper-triangular, and $\B(\Theta,\xi)$ is symmetric. Here $p$ is the dimension of $\Theta$, and $q$ is the dimension of $\xi$. \textbf{Proof:} See the supplementary material.
\end{lemma}

\subsection{Regression Calibration}
Regression calibration using the BLUP technique functions almost equivalently in the case of the generalized error models as in a standard error model. We model $\widehat{X}_i = \widehat{E}[X_i|X_i^*, Z_i]$ as a linear function and then estimate $\widehat{\Theta}_\text{RC}$ by solving $0 = U_n(Y,\widehat{X},Z; \widehat{\Theta}_\text{RC})$. When assuming an additive measurement error model with replicate observations, we take the mean of the replicates as $X_i^*$. In the case of non-identically distributed measurements, it is unlikely that this will be the most efficient combination. Intuitively, measurements with lower variance ought to contribute more to the proxy measure than those with higher variance. We define $X_i^* = \sum_{j=1}^{\kappa_i} \alpha_jX_{ij}^*$, for some set of weights $\{\alpha_j\}$ such that $\sum_{j=1}^{\kappa_i} \alpha_j = 1$, with $\alpha_j \geq 0$ for $j=1,\hdots,\kappa_i$. If we have $\kappa_i = k$, $\alpha_j = \frac{1}{k}$, and $J_{01} = \{1,\hdots,k\}$, then as previously discussed, the estimators become mathematically equivalent to the standard estimators \citep{CarrollBook}. 

It is worth considering, however, that when the repeated measurements are not identically distributed, the parameters in the standard implementation can be hard to interpret. In particular, the estimated measurement error variance corresponds to an estimate of the average measurement error variance across the repeated measurements, not necessarily to any of the specific measurements themselves. In this way, even if the estimate of the average variance is correct, it is not possible to conclude that this estimate corresponds to the error variance for any of the individual error-prone measurements. Additionally, if replicates are incomplete, we have discussed that the standard regression calibration technique will no longer suffice. In addition to errors that this incomplete replication may cause during estimation and inference, it may also cause issues for future investigations. For instance, in the precision medicine setting, analysts may wish to allow the error-prone variable to impact a future treatment recommendation. In the modified framework we can proceed with the correction using any subset of the measurements, allowing for patients outside the sample to be accommodated. In the standard procedure, this is not possible as, generally $\Sigma_{UU} \neq M_j$. 

A final concern is that the standard procedure forces the weights chosen to be $\frac{1}{k}$, which, are unlikely to be optimal. We advocate for adding $\{\alpha_j\}$ as parameters to the BLUP directly, and treat them as a parameter minimizing the MSE between $\widehat{X}_i$ and $X_i$. This provides the set of optimal weights in the same sense that the BLUP provides the optimal set of $\widehat{X}_i$. It will not be possible, in general, to derive a closed form expression for the set of weights and it will instead require numerical optimization, alongside the BLUP parameter estimates.

The BLUP is only a consistent estimator for the conditional expectation, $E[X_i|X_i^*, Z_i]$, when the conditional expectation is linear in the conditioning variables. In the supplementary material, Lemma \ref{lemma::conditional_mean} is provided as a generalization of Lemma A.1 from \citet{Carroll1990}, which uses their notation for matrix derivatives and the trace operator. Using this Lemma, we can characterize the linearity of the BLUP. For notation, the inverse of $v = (v_1, \hdots, v_p)'$ is given by $v^{-1} = (\frac{1}{v_1}, \hdots, \frac{1}{v_p})'$.
\begin{theorem}[General Form of Conditional Means]\label{thm::conditional_means}
    Under the generalized error models presented, assuming that $E[U_i|X_i] = 0$, we have that \begin{equation}E[X_i|X_i^*] = \eta_1^{-1}\left[X_i^* - \eta_0 + \delta^2\left\{\Tr\left(\frac{\partial}{\partial x}\Omega(x)\right) + \Omega(x)\frac{f_{X^*}'(x)}{f_{X^*}(x)}\right\}_{x=X_i^*}\right] + O_p(\delta^3),\label{eq::conditional_mean_add} \end{equation} when $X_i^* = \eta_0 + \eta_1X_i + \delta U_i$ and \begin{align}
        \begin{split}
            E[X_i|X_i^*] &= \eta_1^{-1}\left[1 + \delta^2\left[2\cdot\text{\normalfont diag}\left(\Omega(x)\right) + \right.\right.\\
            &\left.\left.x\circ\left\{\Tr\left(\frac{\partial\Omega(x)}{\partial x}\right) + \Omega(x)\frac{f_{X^*}'(x+\eta_0)}{f_{X^*}(x+\eta_0)}\right\}\right]_{x=X_i^*-\eta_0}\right]\left(X_i^* - \eta_0\right) + O_p(\delta^3)
        \end{split}\label{eq::conditional_mean_mult},
    \end{align} when $X_i^* = \eta_0 + \eta_1X_i(1 + \delta U_i)$. \textbf{Proof:} See Supplementary Appendix A.
\end{theorem}

The term $\frac{f_{X^*}'(x)}{f_{X^*}(x)}$ is linear in $x$ if and only if $X_i^* \sim N(\mu, \sigma^2)$ \citep{Carroll1990}. Since we are conditioning on $X_i^*$, we can exclude values of this ratio which are unobservable almost surely. As a result, domain indicators can be dropped. Then, for the case of additive measurement error, the conditional mean (\ref{eq::conditional_mean_add}) will be approximately linear if either $\Omega(X_i^*)$ is linear and $\frac{f_{X^*}'(x)}{f_{X^*}(x)}$ is constant, or if $\Omega(X_i^*)$ is constant and $\frac{f_{X^*}'(x)}{f_{X^*}(x)}$ is linear. Linearity in the multiplicative mean (\ref{eq::conditional_mean_mult}) is more restrictive. Here, due to the additional multiplicative $X_i^*$ term, we need both $\text{diag}(\Omega(X_i^*-\eta_0))$ to be constant and $0 = \Tr\left(\frac{\partial\Omega(x)}{\partial x}\right) + \Omega(x)\frac{f_{X^*}'(x+\eta_0)}{f_{X^*}(x+\eta_0)}$. If $\Omega(x)$ is constant, then the first term of equation (\ref{eq::conditional_mean_mult}) will be $0$. The second term is $0$ only if $\cov(U_i|X_i) = 0$ or $f_{X^*}(x)$ is constant. As a result, it is sufficient to have $X_i^* \sim \text{Unif}(\cdot)$, and for $\Omega(x)$ to be constant. This illustrates the caveats with applying this method to multiplicative errors. In many situations, the linear approximation for the additive case will be sufficiently good. However, in the multiplicative case the expectation will be non-linear under most assumed models. 

Consider the case when no $Z_i$ is measured, and $X_i$ is a scalar. In order for $E[X_i|X_{ij}^*]$ to be linear, additive models require $E[U_{ij}|X_{ij}^*]$ to be linear, and multiplicative models require $E[(1+\delta_lU_{il})^{-1}|X_{il}^*]$ to be constant. If we have two observations from the generalized error model, $X_{i1}^*$ and $X_{i2}^*$, then direct calculations show that $E[X_{i1}^*|X_{i2}^*]$ will be linear under the same conditions that render $E[X_i|X_{i2}^*]$ linear. This applies symmetrically to $E[X_{i2}^*|X_{i1}^*]$. Checking the goodness of fit of a linear model between any two proxies in turn checks the adequacy of a linear approximation. This also highlights the relationship between the proposed methodology and the instrumental variable approaches, which are based on regressing a measurement of the truth on an instrument \citep{CarrollBook}. These results justify both the theoretical conditions under which a linear model is warranted, and a technique for checking whether linearity holds approximately. The following result establishes that the modified regression calibration procedure produces asymptotically normal estimators.
\begin{theorem}[Asymptotic Normality of Regression Calibration]\label{thm::asymptotic_normality_rc}
    Under standard regularity conditions, the estimated parameters using the regression calibration correction, $\widehat{\Theta}_\text{RC}$ are consistent for the parameters $\Theta_\text{RC}$, and are asymptotically normally distributed, such that, as $n\to\infty$, $\sqrt{n}\left(\widehat{\Theta}_\text{RC} - \Theta_\text{RC}\right) \stackrel{d}{\longto} N\left(\mathbf{0},\Sigma_\text{RC}\right)$,  where $\Sigma_\text{RC} = Q\A_\text{RC}^{-1}\B_\text{RC}\A_\text{RC}^{-'}Q'$, for matrices analogous to those in Lemma \ref{lemma::general_asymptotic_distribution}. \textbf{Proof:} See Supplementary Appendix A.
\end{theorem}

Importantly, this result shows asymptotic normality, not around the true value, $\Theta_0$, but around $\Theta_\text{RC}$, the solution to $0 = U(Y,\widehat{X},Z,\Theta_{RC})$. Under regularity conditions, $\Theta_\text{RC}$ is the probability limit of $\widehat{\Theta}_\text{RC}$. The asymptotic bias of the regression calibration correction is thus determined by the difference between $\Theta_0$ and $\Theta_\text{RC}$, with consistency achieved when $\Theta_0 = \Theta_\text{RC}$. The previously discussed results regarding consistency, and approximate consistency, of regression calibration methods generally apply to the modified technique, under the caveat that these are derived when $\widehat{X}_i \stackrel{p}{\longto} E[X_i|X_i^*,Z_i]$.

\subsection{Simulation Extrapolation}\label{subsec::SIMEX}
To modify the SIMEX corrections, it is insufficient to add pseudo-errors with variance $\lambda\delta$, as $\delta$ will not be constant across $j$. Instead, we rely on $M_j$ and $M_j(X)$ to motivate the modified versions of SIMEX. The strategy will be to match the first two moments of $X_i$ and $X^*_{b,i}(\lambda)$, as $\lambda\to-1$. For fixed $\lambda \geq 0$, we take \begin{equation}
    X_{bij}^*(\lambda) = \eta^{-1}_{1j}\circ\left(X_{ij}^* - \eta_{0j} + \sqrt{\lambda}M_j^{1/2}\nu_{bj}\right),\label{eq::modified_simex_form}
\end{equation} where $\nu_{bj}$ is an appropriately sized standard normal random variable, independent of all covariates. Given $X_i$, we find that $E[X_{bij}^*(\lambda)|X_i] = X_i$, and  $\cov\left(\left.X_{bij}^*\right|X_i\right) = (1+\lambda)\eta_{1j}^{(d)-1}M_j(X_i)\eta_{1j}^{(d)-1}$. As a result, $X_{bij}^*(\lambda)$ agrees with $X_i$ up to the second moment, as $\lambda\to-1$.

This modified procedure treats $M_j$ as the variance of the pseudo-random variable that is generated. As a result, our previous discussion regarding when $M_j$ is positive semidefinite apply here. As a result, the SIMEX correction can be applied in the generalized framework only when, either, the measurement error can be safely assumed to be additive, or if the error-prone variable is univariate. As in the standard SIMEX, we do not typically have $M_j$, $\eta_{0j}$, or $\eta_{1j}$ available, and as a result we will estimate them from the proxy observations. In the standard measurement error setting, if homoscedacticity is assumed, then SIMEX progresses using $\overline{X}^*$ and $\widehat{\Sigma}_{UU}$. As discussed with regression calibration, if $\kappa_i = k$, the standard SIMEX applies to non-iid data, with the same caveats. The empirical SIMEX relies more heavily on the iid property, and while it is not strictly necessary, it is not readily facilitated in the more general case. As a result, we continue to focus on homoscedastic errors.

We consider two ways of combining the replicate measurements based on equation (\ref{eq::modified_simex_form}): averaging estimates of $\Theta_0$ or averaging the proxies. To average estimates, we find $\widehat{\Theta}_\text{SIMEX}^{(j)}$ for each $j=1,\hdots,k$, and combine these $k$ estimates. Alternatively, we could use (\ref{eq::modified_simex_form}), where in place of $X_{ij}^*$ we take $X_i^* = \sum_{j=1}^k \alpha_jX_{ij}^*$ for some set of weights $\{\alpha_j\}$.

Just as with regression calibration, the SIMEX correction is only approximately consistent. The quality of the approximation is determined by (1) the quality of the extrapolant, and (2) how well the matching of the first two moments of $X_{bij}^*(-1)$ and $X_i$ approximates the matching of their distributions. As a result, if the extrapolant is correctly specified and, for instance, $X_{bij}^*(\lambda)$ is normally distributed, then SIMEX is consistent. $\widehat{\Theta}_\text{SIMEX}$ will be consistent for $\lim_{\lambda\to-1} \G\left\{X_{bij}^*(\lambda)\right\}$, which we call $\Theta_\text{SIMEX}$, and will generally be asymptotically normal.

\begin{theorem}[Asymptotic Normality of SIMEX]\label{thm::simex_asymptotic_normality}
    Under standard regularity conditions, the estimated parameters using the SIMEX correction, $\widehat{\Theta}_\text{SIMEX}$ are consistent for the parameters $\Theta_\text{SIMEX}$, and as $n\to\infty$, $\sqrt{n}\left(\widehat{\Theta}_\text{SIMEX} - \Theta_\text{SIMEX}\right) \stackrel{d}{\longto} N(\mathbf{0},\Sigma_\text{SIMEX})$, where $\Sigma_\text{SIMEX}$ is estimable through sandwich estimation techniques. \textbf{Proof:} See Supplementary Appendix A.
\end{theorem}

\section{Simulation Studies}\label{sec::simulations}
To investigate the behavior of the proposed methods in the generalized measurement error framework, we consider three simulated scenarios, comparing them with the standard implementation of these techniques. We consider a variety of settings for which SIMEX and regression calibration are known to be effective, when the repeated measurements are iid. 

\subsection{Linear Regression Models}
First we consider a linear regression. We take $X = \begin{bmatrix}X_1 & X_2 & X_3\end{bmatrix}$, with $X_1 \sim N(0,1)$, $X_2 \sim N(3, 2)$, and $X_3 \sim N(1,3)$ to be the true covariate vector, where all components are assumed to be independent. The outcome is taken to be $Y = 2 - X_1 + 2X_2 + 0.5X_3 + \epsilon$, where $\epsilon \sim N(0,1)$. We generate three error-prone proxies, $X_1^* = X + \begin{bmatrix}U_{11} & U_{12} & U_{13}\end{bmatrix}$, where $U_{1j} \sim N(0,1)$, $X_2^* = X + \begin{bmatrix} U_{21} & U_{22} & U_{23} \end{bmatrix}$ with $U_{21} \sim N(0,1)$, $U_{22} \sim N(0,4)$ and $U_{23} \sim N(0,3)$, and $X_3^* = X + \begin{bmatrix}U_{31} & U_{32} & U_{33}\end{bmatrix}$ where $U_{31} \sim N(0,2)$, $U_{32} \sim N(0,2)$ and $U_{33} \sim N(0,5)$. We select $50\%$ of $X_2^*$ and $20\%$ of $X_3^*$ to be missing, completely at random. All error terms are generated independently of each other and of all other quantities.

\begin{figure}[h]
    \begin{center}
        \includegraphics[width=0.98\textwidth]{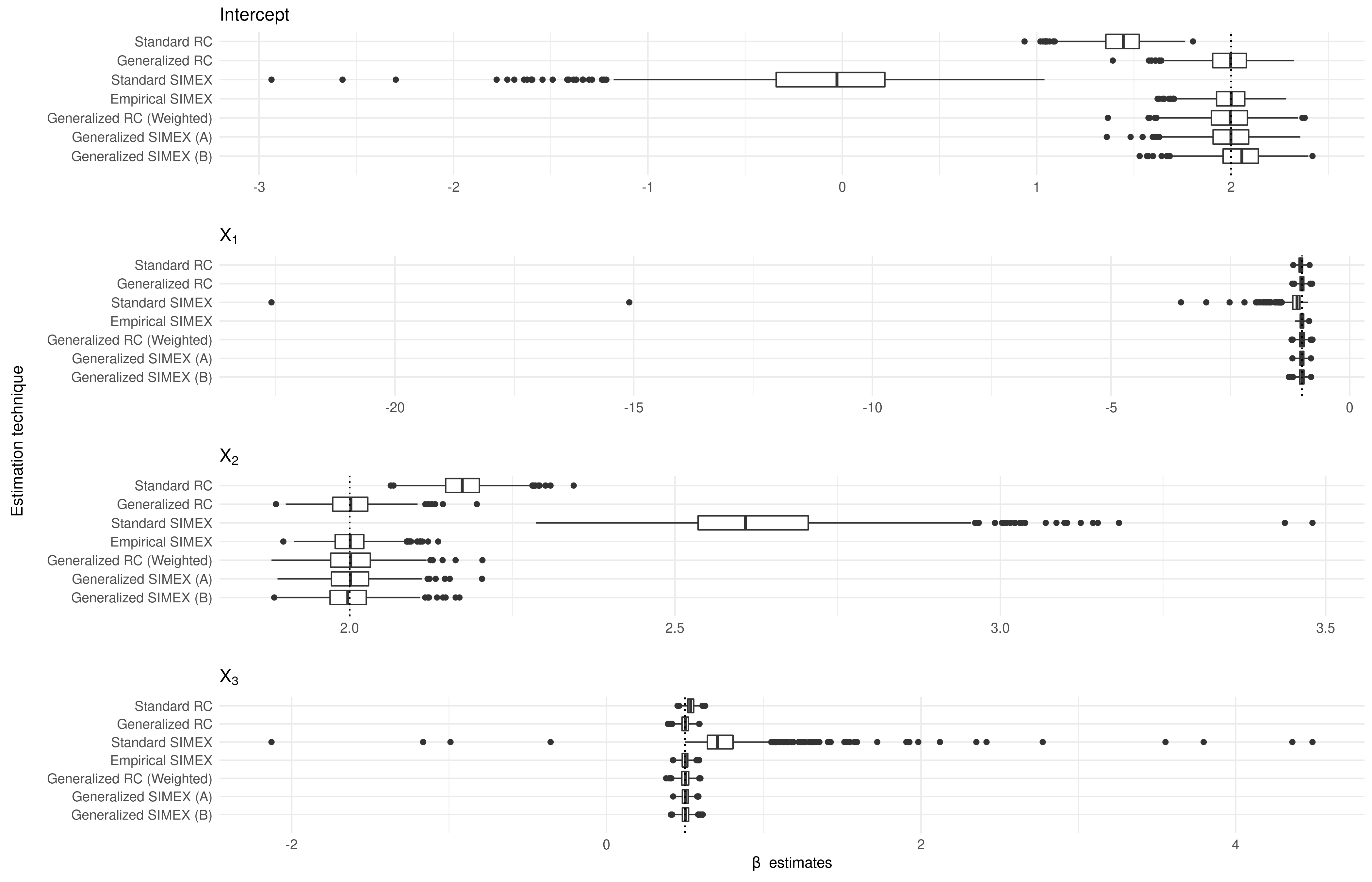}
    \end{center}
    \caption{Simulation results of a linear regression model comparing the generalized and standard techniques. The plots show the intercept and slopes (for $X_1$, $X_2$, and $X_3$) that were estimated in $1000$ ($n=5000$) replicated simulations using (1) standard regression calibration (Standard RC), (2) standard SIMEX, (3) empirical SIMEX, (4) generalized regression calibration using fixed weights as $\frac{1}{\kappa_i}$ (Generalized RC), (5) generalized regression calibration solving for optimal weights (Generalized RC (Weighted)), (6) generalized SIMEX where proxies are combined (Generalized SIMEX (First)), and (7) generalized SIMEX where the estimates are combined (Generalized SIMEX (Second)). The true parameter values are indicated with dotted lines. Outliers are displayed as filled in circles. Note that the $x$-axes are different for each set of box plots.}
    \label{fig::sim1_boxplot}
\end{figure}

We estimate model parameters using (1) standard regression calibration, (2) standard SIMEX, (3) empirical SIMEX, (4) generalized regression calibration using fixed weights as $\frac{1}{\kappa_i}$, (5) generalized regression calibration solving for optimal weights, (6) generalized SIMEX where proxies are combined, and (7) generalized SIMEX where the estimates are combined. These simulations were repeated $1000$ times with a sample size of $5000$. The results for all scenarios are included in Figure \ref{fig::sim1_boxplot}. 

The generalized procedures generally correctly estimate the parameter values, producing similar results. As expected, the standard regression calibration and SIMEX procedures perform poorly.

\subsection{Log-Linear Regression Models}
Next, we consider a log-linear model. We generate $Z \sim \text{Binom}(0.3)$ and $X \sim N(0.02Z, 0.5)$. The outcome is a gamma random variable with $E[Y|X, Z] = \exp\left(2 - 3Z + 2X\right)$. We generate three error-prone proxies, $X_1^* = XV_1$ where $V_1 \sim \text{Unif}(0.7, 1.3)$, $X_2^*,X_3^*\stackrel{iid}{\sim} X + N(0,1)$, and the errors are all independent. For $X_3^*$ we selected $50\%$ of the observations to be missing completely at random. We compare nine different methods, the seven introduced for the linear regression simulations, as well as both the weighted and unweighted regression calibration correction where the correction parameters are computed ignoring $Z$.

\begin{figure}
    \begin{center}
        \includegraphics[width=0.98\textwidth]{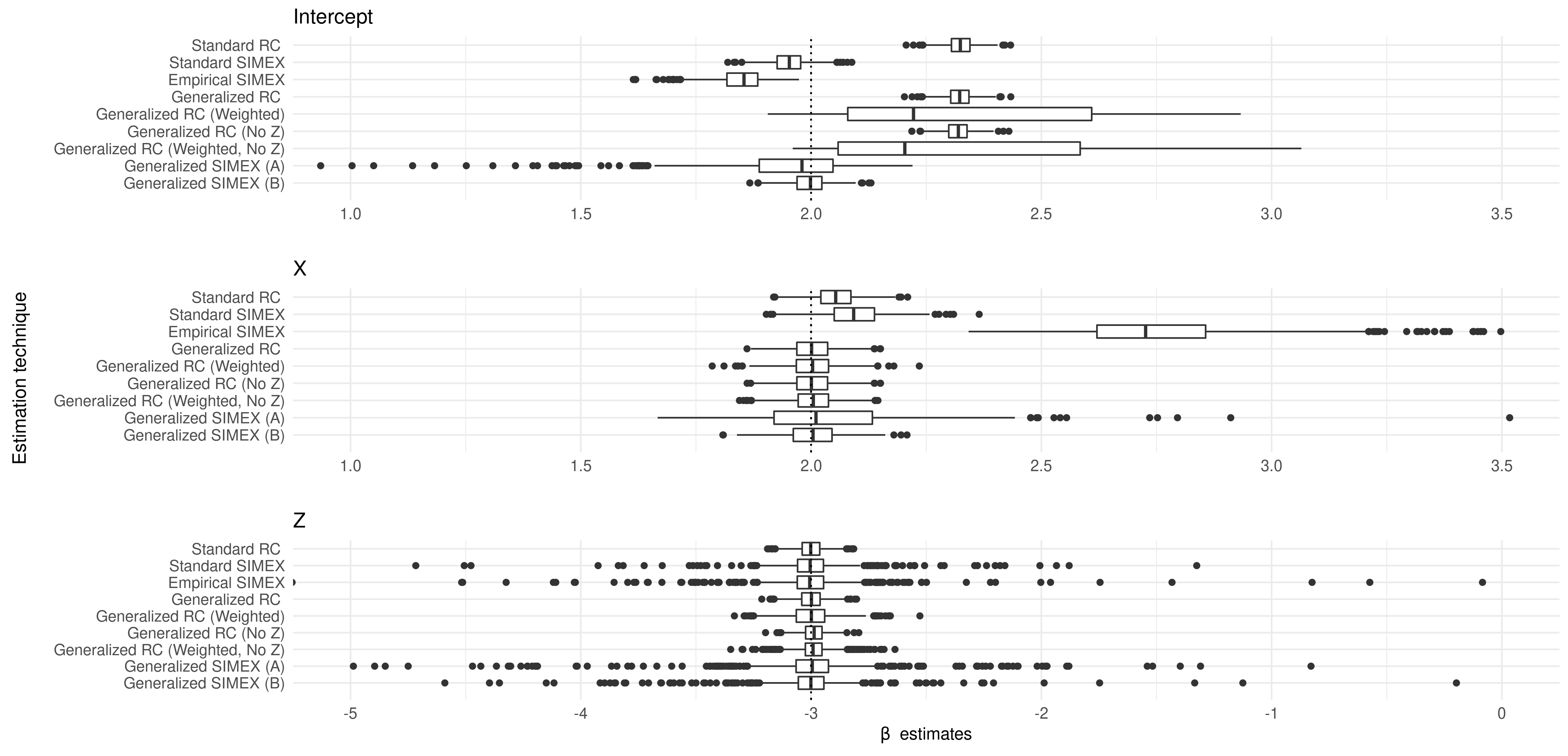}
    \end{center}
    \caption{Simulation results of a gamma, log-linear regression model comparing the generalized and standard techniques. The plots show estimates for the intercept and slopes (for $X$ and $Z$) that were estimated in $1000$ ($n=5000$) replicated simulations using (1) standard regression calibration (Standard RC), (2) standard SIMEX, (3) empirical SIMEX, (4) generalized regression calibration using fixed weights as $\frac{1}{\kappa_i}$ (Generalized RC), (5) generalized regression calibration solving for optimal weights (Generalized RC (Weighted)), (6) generalized regression calibration using fixed weights as $\frac{1}{\kappa_i}$ without using $Z$ for the parameter estimation (Generalized RC (No Z)), (7) generalized regression calibration solving for optimal weights without using $Z$ for parameter estimation (Generalized RC (Weighted, No Z)), (8) generalized SIMEX where proxies are combined first (Generalized SIMEX (First)), and (9) generalized SIMEX where the estimates are combined (Generalized SIMEX (Second)). The true parameter values are indicated with dotted lines. Outliers are displayed as filled in circles. Note that the X axes are different for each set of box plots.}
    \label{fig::sim2_boxplot}
\end{figure}

The results are summarized in Figure \ref{fig::sim2_boxplot}. The generalized regression calibration methods perform well for the slope parameters, and the generalized SIMEX methods perform well for all three parameters. We also emphasize that these results are based on a multiplicative error model, showcasing the capacity of both methods to function in this scenario.

\subsection{Logistic Regression Models}
Finally, we consider a logistic regression model. We take the true covariate $X \sim N(3,1)$, with $Y\sim\text{Binom}(\text{expit}(0.5 - 0.5X))$, where $\text{expit}(u) = \{1 + \exp(-u)\}^{-1}$. We generate three error-prone proxies where $X_1^*, X_2^* \stackrel{iid}{\sim} X + N(0,1)$, and $X_3^* = 0.5 + 0.5X + U_3$ where $U_3 \sim \text{Unif}(-0.5, 0.5)$. We select $80\%$ of $X_2^*$ to be missing. In these simulations we compare the results of the generalized estimators, using either all of $\{X_1^*, X_2^*, X_3^*\}$ or only the iid replicates $\{X_1^*, X_2^*\}$ for the corrections, labeled ``all'' and ``IID'' respectively. We differentiate between the weighted generalized regression calibration and the standard version, as well as the SIMEX estimator that averages proxies versus the one which averages estimates.

\begin{figure}
    \begin{center}
    \includegraphics[width=0.98\textwidth]{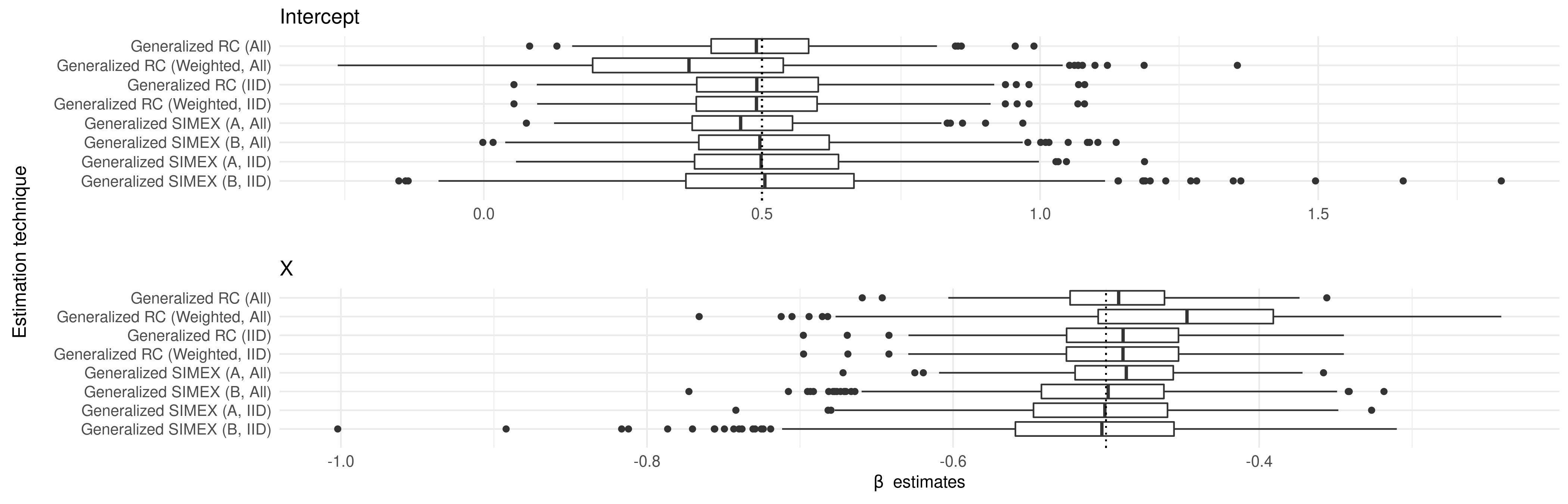}
    \end{center}
    \caption{Simulation results of a logistic regression model comparing the proposed generalized techniques. The results are for $1000$ ($n=5000$) replicates of the simulation using (1) generalized regression calibration using all proxies (Generalized RC (All)), (2) generalized regression calibration with optimal weights, using all proxies (Generalized RC (Weighted, All)), (3) generalized regression calibration using only the replicates (Generalized RC (IID)), (4) generalized regression calibration with optimal weights, using only the replicates (Generalized RC (Weighted, IID)), (6) generalized SIMEX where the proxies are averaged, using all proxies (Generalized SIMEX (First, All)), (7) generalized SIMEX where the estimates are averaged, using all proxies (Generalized SIMEX (Second, All)), (8) generalized SIMEX where the proxies are averaged, using only replicates (Generalized SIMEX (First, IID)), and (9) generalized SIMEX where the estimates are averaged, using only replicates (Generalized SIMEX (Second, IID)). The estimated parameter values for the intercept and the slope across the different methods are given. The true values are indicated using a dotted line. Outliers are displayed as filled in circles. Note that the X axes are different for each set of box plots.}
    \label{fig::sim3A}
\end{figure}

\begin{figure}
    \begin{center}
    \includegraphics[width=0.98\textwidth]{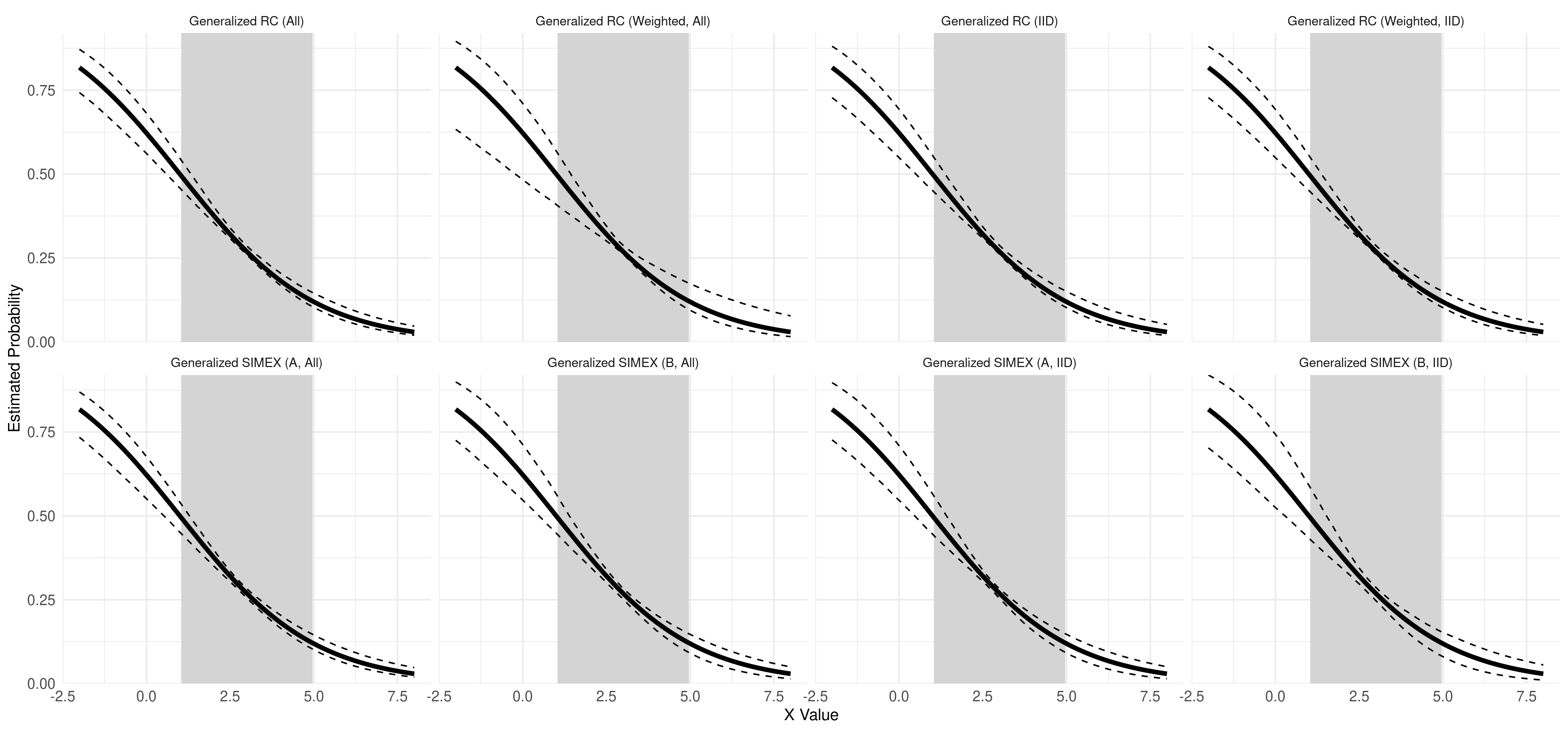}
    \end{center}
    \caption{Simulation results of a logistic regression model comparing the proposed generalized techniques. The results are for $1000$ ($n=5000$) replicates of the simulation using (1) generalized regression calibration using all proxies (Generalized RC (All)), (2) generalized regression calibration with optimal weights, using all proxies (Generalized RC (Weighted, All)), (3) generalized regression calibration using only the replicates (Generalized RC (IID)), (4) generalized regression calibration with optimal weights, using only the replicates (Generalized RC (Weighted, IID)), (6) generalized SIMEX where the proxies are averaged, using all proxies (Generalized SIMEX (First, All)), (7) generalized SIMEX where the estimates are averaged, using all proxies (Generalized SIMEX (Second, All)), (8) generalized SIMEX where the proxies are averaged, using only replicates (Generalized SIMEX (First, IID)), and (9) generalized SIMEX where the estimates are averaged, using only replicates (Generalized SIMEX (Second, IID)). The estimated $95\%$ prediction intervals for the estimated probabilities (given by the dotted lines) around the true probabilities (given by the solid line), across various values for $X$ are given. The shaded regions indicate the $95\%$ central values for $X$, indicating the most likely values for the covariate to take.}
    \label{fig::sim3B}
\end{figure}

In Figure \ref{fig::sim3A} we observe the results of the parameter estimates and in Figure \ref{fig::sim3B} we observe the results of the estimated probabilities. These demonstrate the bias reduction and effective probability estimates of both techniques in logistic regression. Moreover, these simulations demonstrate how biased (both using $\eta_0$ and $\eta_1$) proxies can stabilize estimators.


\section{Extensions to Other Methodologies}\label{sec::extensions}
Moment reconstruction \citep{MomentReconstruction} is a method that was developed in a spirit similar to regression calibration, but which sought to overcome some of its shortcomings. In particular, analysis is conducted using $\widehat{X}_\text{MR,i}$ substituted for $X_i$, where $\widehat{X}_\text{MR,i}$ is an estimated version of $X_\text{MR,i}$ selected such that the joint distribution $(Y_i,X_\text{MR,i}) \sim (Y_i,X_i)$. Unlike regression calibration, moment reconstruction involves solving for the forms of parameters based on the assumed underlying model. We present the results of moment reconstruction in a logistic regression, a case where the moment reconstruction estimators are consistent whereas the regression calibration corrections are not. The primary motivation for this presentation is to demonstrate how the identification of parameters as in Section \ref{subsec::parameter_identification}, and the related results, can be extended to some exact correction methods, using problem-specific derivations.

Assume that $X_i|Y_i=y \sim N(\mu + y\Delta, \Sigma_{XX})$. Then, for each observation, moment reconstruction forms $\widehat{X}_\text{MR,i}(X_i^*, Y_i) = \frac{1}{\eta_{1\cdot}}\left[\left\{E[X_i^*|Y_i](I-\widetilde{G}(Y_i)) + X_i^*\widetilde{G}(Y_i)\right\}-\eta_{0\cdot}\right]$, where $\widetilde{G}(Y_i) = \eta_{1\cdot}G(Y_i)$, $\eta_{l\cdot} = \sum_{j=1}^k\alpha_j\eta_{lj}$ for $l=1,2$, and $G(Y_i) = \cov(X_i^*|Y_i)^{-\frac{1}{2}}\cov(X_i|Y_i)^{1/2}$. This results in $\widehat{X}_{MR,i}$ having the same first two conditional moments, given $Y_i$, as $X_i$ does. This setup readily presents M-estimators, extending the quantities in Section \ref{sec::generalized_methods}. These are presented explicitly in Supplementary Appendix B.

Denoting the probability $P(Y=1|X=x) = \text{expit}(\beta_0 + \beta_1x)$, the logistic regression estimators for $\beta_0$ and $\beta_1$ are $\widehat{\beta}_0$ and $\widehat{\beta}_1$, with these estimators simultaneously solving $0 = n^{-1}\sum_{i=1}^n \left\{y_i - \text{expit}\left(\widehat{\beta}_0 + \widehat{\beta}_1x_i\right)\right\}$ and $0 = n^{-1}\sum_{i=1}^n \left\{y_i - \text{expit}\left(\widehat{\beta}_0 + \widehat{\alpha}_1x_i\right)\right\}x_i$. The moment reconstruction procedure replaces $x_i$ in the above estimating equations with, \begin{align*} &\left(\sum_{j=1}^k \alpha_j\eta_{1j}\right)^{-1}\left\{\left[y_i\Theta_1 + (1-y_i)\Theta_2\right]\left[1-\left(\sum_{j=1}^k\alpha_j\eta_{1j}\right)\left(\frac{\Sigma_{XX}}{y_i\Theta_3 + (1-y_i)\Theta_4}\right)^{1/2}\right]\right.\\
    &\left.+\left(\sum_{j=1}^k\alpha_jx_{ij}^*\right)\left(\sum_{j=1}^k\alpha_j\eta_{1j}\right)\left(\frac{\Sigma_{XX}}{y_i\Theta_3 + (1-y_i)\Theta_4}\right)^{1/2} - \sum_{j=1}^k\alpha_j\eta_{0j}\right\},
\end{align*} denoted $\widehat{x}_i$. This expression can be inserted into the M-estimators for $\widehat{\beta}$, and stacked with the parameter M-estimators, allowing for the derivation of an asymptotic distribution. Due to normality, the distributions $(\widehat{X}_\text{MR,i},Y_i)$ and $(X_i,Y_i)$ are equivalent, and so this estimator will be consistent and asymptotically normal for the true parameter values. Further, the solutions to the M-estimators regarding the parameters in $g_i(\cdot)$ and the estimators for $\Theta_{\cdot}$ are often expressible in closed form, and are functionally independent of $\beta_{\cdot}$. As a result, they can be solved for first and used to compute $\widehat{x}_i$, before performing a standard logistic regression.

In order to implement this in practice, or compute an explicit expression for the asymptotic distribution using Lemma \ref{lemma::general_asymptotic_distribution}, we must make concrete assumptions regarding the replicate measurements that are available and the values of $\{\alpha_j\}$. These data must conform to the identifiability conditions for $g$. 

\section{Framingham Heart Study}\label{sec::analysis}
We now apply the generalized methods to data from the Framingham Heart Study. Our analysis is motivated by \citet{CarrollBook}, where the authors use a logistic regression model to estimate the impact of, age, smoking status, serum cholesterol, and long-term SBP on the likelihood of developing CHD. Our analysis follows a different susbet from the FHS. Our subset is not restricted to male participants, and so we use sex as an explanatory factor as well. We use a different subset of data from the same study which is made available as a teaching dataset, by the NHLBI \citep{TeachingData}.

Our analysis follows 2876 individuals, aged 32--69, across three separate examinations. We take the patients' sex, age, and smoking status to be error-free, and assume that the serum cholesterol levels and systolic blood pressure are prone to error. Following \citet{Cornfield} and \citet{CarrollFHS} we transform the blood pressure measurements to be included in the model as $\log\left(\text{SBP} - 50\right)$ and the cholesterol measurements to be included as $\log\left(\text{Cholesterol}\right)$. These data are subject to incomplete replication. Of the 2876 total participants, systolic blood pressure measurements were available for all patients at the first visit, but missing for 153, and 390 patients at visits two, and three respectively. For cholesterol, at visits one, two, and three, there are 26, 256, and 538 patients without replicate measurements respectively. Considering only those with the replicate measurements taken, at the first visit the mean (transformed) SBP was 4.329 and the mean (transformed) cholesterol was 5.437, with observed variances of 0.052 and 0.033, respectively. This is compared to means (variances) of SBP and cholesterol at the second visit of 4.389 (0.054) and 5.503 (0.030), and at the third visit 4.440 (0.057) and 5.456 (0.033), respectively.

We compare several different analyses, all of which use the main effects model in a standard logistic regression. We consider a naive analysis, which takes the mean response from the visits for both cholesterol and blood pressure as the explanatory factors, a standard regression calibration analysis which implicitly assumes that the replicate measurements are iid, and several scenarios for the generalized procedures presented. We consider different assumptions for $J_0$, the proxies which have $\eta_{0j} = 0$, and $J_1$, those with $\eta_{1j} = 1$. For regression calibration we consider four scenarios, two with $J_0 = \{1,2,3\}$, where $J_1 = \{1,2,3\}$ or $J_1 = \{1,2\}$, in addition to two with $J_0 = \{2\}$, with $J_1 = \{1,3\}$ or $J_1 = \{2,3\}$. We consider two SIMEX analyses, one with $J_0 = J_1 = \{1,2,3\}$, and one with $J_0 = 2$ and $J_1 = \{1,3\}$. The SIMEX procedures are restricted in their consideration due, in part, to the concerns regarding the validity of $M_j$ as a variance matrix. Many plausible settings lead to singular matrices as estimates for $M_j$, which in turn rules out the use of the modified SIMEX under those assumptions. The SIMEX procedures used a quadratic extrapolant for both the SBP and the cholesterol terms.

The results of these analyses are displayed in Table \ref{tb::fram_results}, where the slope parameter estimates for the transformed systolic blood pressure and the transformed cholesterol are presented, along with $95\%$ bootstrapped confidence intervals. The bootstrap confidence intervals are derived from $1000$ bootstrap replicates in each scenario. Across the various different sets of assumptions, we observe some variability in the point estimates for both factors, with more substantial variability in the cholesterol measurements. While none of the methods find the effect of cholesterol to be significant at a 95\% confidence level, the implied level of significance varies across the scenarios. 

\begin{table}[ht]
    \caption{Estimated slope parameter for the SBP and cholesterol  terms, in the FHS, comparing the generalized regression calibration and SIMEX methodologies to a naive analysis and standard regression calibration. The point estimates and $95\%$ confidence interval are shown, where the intervals are based on a bias corrected bootstrap procedure with $1000$ bootstrap replicates.}
    \centering
    \begin{tabular}{lrr}
      \hline
      & \multicolumn{1}{c}{Blood Pressure} & \multicolumn{1}{c}{Cholesterol} \\
      \hline
      Naive & 2.2502 (1.6956, 2.8369) & 0.6698 (-0.0826, 1.5747) \\
      Standard Regression Calibration & 2.8113 (2.1037, 3.5911) & 0.7534 (-0.1773, 1.8657) \\[0.1cm]
      \multicolumn{2}{l}{Generalized Regression Calibration} & \\
      $J_0 = J_1 = \{1,2,3\}$ & 2.6876 (2.0047, 3.4174) & 0.7234 (-0.1712, 1.7901) \\
      $J_0 = \{1,2,3\}$; $J_1 = \{1,2\}$ & 2.6732 (1.9924, 3.4124) & 0.9345 (-0.1375, 2.2072) \\
      $J_0 = 2$; $J_1 = \{1, 3\}$ & 2.6352 (1.9171, 3.4150) & 0.7317 (-0.1681, 1.8083) \\
      $J_0 = 2$; $J_1 = \{2, 3\}$ & 2.7852 (2.0970, 3.5496) & 0.3467 (-0.2003, 1.2763) \\[0.1cm]
      \multicolumn{2}{l}{Generalized Simulation Extrapolation} & \\
      $J_0 = J_1 = \{1,2,3\}$ & 2.6741 (1.6262, 5.8917) & 0.9996 (-0.4762, 15.1816) \\
      $J_0 = 2$; $J_1 = \{1, 3\}$  & 2.0963 (1.1682, 6.4064) & 0.5667 (-2.4536, 3.7640) \\
       \hline
    \end{tabular}
    \label{tb::fram_results}
\end{table}

\section{Discussion}\label{sec::discussion}
Regression calibration and simulation extrapolation are frequently used techniques designed to improve analyses where covariates are measured with error. These techniques traditionally assume that the analyst has access to validation data or identically distributed repeat measurements in order to facilitate the correction. Validation data are often impossible to collect, and while there may be replicate measurements available, the assumption that these measurements are identically distributed may be violated. This is a concern given the broad appeal of these techniques. We have shown how regression calibration and SIMEX can be used when the repeated measurements available are not identically distributed, while maintaining their comparatively simple implementation.

Regression calibration and SIMEX are most commonly implemented only for the classical additive model. Our corrections are presented for both a biased additive model, as well as a biased multiplicative structure. We echo the common sentiment that transformations to additivity are preferable when available. However, our methodology presents the flexibility of accommodating both types of error without the analyst needing to identify which is present. To ensure confidence in the applicability of these methods, we discussed how standard regression diagnostics can be used to determine whether the corrections are reasonable to apply. 

The behavior of our corrections are shown, across popular outcome models, to perform satisfactorily when these standard assumptions are violated. In addition to simulation studies, we have presented asymptotic results justifying their utility. Under regularity conditions the estimators for the parameters required to make the corrections are consistent and asymptotically normal. Using these estimated parameters in regression calibration or SIMEX leads to approximately consistent results. If exact consistency is required, we have shown how this generalized error model can be accommodated in other error correction techniques as well.

We applied our methods to The Framingham Heart Study, which has been shown to violate the iid replicate assumption. This application demonstrates a setting where the violation of the iid assumption may have a more substantive impact on conclusions drawn from an analysis. The data we analyzed contains data from the Framingham study as collected, but due to anonymization techniques, it is not appropriate for drawing scientific conclusions. Instead, our analysis uses these data to demonstrate the utility of the methods presented, noting that a similar analysis would need to be done on the complete data, with input from subject matter experts, in order to draw health-related conclusions. Our results do suggest that a consideration of the underlying assumptions regarding the structure of available replicates is important.

There are scenarios, as was seen in the FHS application, where the full generality of the presented model leads to large variances and unstable estimation in the modified SIMEX procedure. We have seen that, while multiplicative errors can stabilize estimation, the use of these measurements is quite limited, and we recommend transformations to additivity where possible. Further, to render these models identifiable, some unbiased measurements are required. It is also important to note that, when replication is complete, the standard methods will accommodate the modified structure without change. Still, assuming that available replicates are identically distributed is an unnecessary restriction on the methods, one which can confuse the interpretation of parameter estimates and lead to inconsistent results where replicates are incomplete. Even if only classical additive error is considered, the techniques presented here showcase that these common corrections can be easily adapted to allow for varying error distributions.

Our corrections are intended to be used in the same contexts as the standard SIMEX and regression calibration procedures. They are no more complicated conceptually, and can be implemented using standard software. The modest additional computational burden of our corrections affords greater flexibility in the data that can be analyzed. There is opportunity to explore the relation of these techniques to instrumental methods, and to explore similar frameworks in exactly consistent corrections.


\appendix
\section{Supplementary Appendix A}
\begin{lemma}
    \label{lemma::conditional_mean}
    Assume that $V_1$, $V_2$, and $V_3$ are random vectors, and that $\delta > 0$ is a constant scalar. Take $E[V_3|V_2] = 0$, and denote $\cov(V_3|V_2=v) = \Omega(v)$. Assume that $E[V_3|V_1]$ and $\cov(V_3|V_1)$ are three-times differentiable functions of $\delta$, a.s. Then \begin{enumerate}[label=(\alph*)]
        \item If $V_1 = V_2 + \delta V_3$, then \begin{equation}
            E[V_3|V_1] = -\delta\left[\Tr\left\{\frac{\partial}{\partial v_1}\Omega(v_1)\right\} + \Omega(v_1)\frac{f_{V_1}'(v_1)}{f_{V_1}(v_1)}\right]_{v_1=V_1} + O_p(\delta^2),
        \end{equation} and \begin{equation}
            \cov(V_3|V_1) = \Omega(V_1) + O_p(\delta).
        \end{equation}
        \item If $V_1 = V_2(\mathbf{1} + \delta V_3)$, then \begin{equation}
            E[V_3|V_1] = -\delta\left[\text{\normalfont diag}\left\{\Omega(v_1)\right\} + v_1\circ\Tr\left\{\frac{\partial}{\partial v_1}\Omega(v_1)\right\} + v_1\circ\Omega(v_1)\frac{f_{V_1}'(v_1)}{f_{V_1}(v_1)}\right]_{v_1=V_1} + O_p(\delta^2),
        \end{equation} and \begin{equation}
            \cov(V_3|V_1) = \Omega(V_1) + O_p(\delta).
        \end{equation}
    \end{enumerate}
\end{lemma}
\begin{proof}[Proof of Lemma \ref{lemma::m_estimator_parameters}]
    Note that, for all components of $\xi$, the provided estimators are independent of $i$. For any such estimator, say given by $\zeta = g(X)$, we can write this as $0 = n^{-1}\sum_{i=1}^n\left\{g(X) - \zeta\right\}$. Our estimators are functions of the $k^2$ parameters given by $\mu_j \equiv E[X_j^*]$ and $\Sigma_{X_j^*X_l^*} = \cov(X_j^*, X_l^*)$, which are to be estimated using ``standard techniques'' from the data. 

    We can estimate these parameters jointly using an M-estimator. Taking $\{C_j\}_{j}$ to denote stacking over $j$ (for instance, $\{X_j^*\}_j = \begin{bmatrix}X_1^* & X_2^* & \cdots & X_k^*\end{bmatrix}'$), and defining $\zeta = (\{\zeta_{1(j)}\}_j, \{\zeta_{2(j,l)}\}_{j,l})'$ to be the parameters $(\{\mu_j\}_j, \{\Sigma_{X_j^*X_l^*}\}_{j,l})'$, then we can take an estimate for $\zeta$ as $\widehat{\zeta}$ which solves 
    \[0 = n^{-1}\sum_{i=1}^n\begin{bmatrix}
        \left\{X_{ij}^* - \zeta_{1(j)}\right\}_{j} \\
        \left\{(X_{ij}^* - \zeta_{1(j)})(X_{il}^* - \zeta_{1(l)}) - \zeta_{2(j,l)}\right\}_{j, l}
    \end{bmatrix}.\] Then, since all of the remaining estimators are independent of $i$, we can use the above form as well as the M-estimators for the $k^2$ parameters to form a joint M-estimator by stacking these together, inserting the relevant $\zeta$ components into the equations. 

    In the event that $X_j^*$ is $p\times1$, with $p\neq 1$, then these can be modified by careful stacking and watching of dimensions. For instance, considering $0 = n^{-1}\sum_{i=1}^n X_{ij}^{*'} - \zeta_{1(j)}$, would give an M-estimator for $\mu_j'$. Stacking these as before, and then modifiying the covariance estimators to be $0 = n^{-1}\sum_{i=1}^n (X_{ij}^* - \zeta_{1(j)}')(X_{il}^* - \zeta_{1(l)}')' - \zeta_{2(j,l)}$, would create an M-estimator as a $\{k + pk(k-1)\}\times p$ matrix. Typically we want this to be an $m\times 1$ function, and so we can vectorize the matrix to achieve that goal. That is, we could solve for $\widehat{\zeta}$ written as a $\left\{(k + pk(k-1))p\right\}\times 1$ vector, using \[0 = n^{-1}\sum_{i=1}^n\text{vec}\left(\begin{bmatrix}
        \left\{X_{ij}^* - \zeta_{1(j)}\right\}_{j} \\
        \left\{(X_{ij}^* - \zeta_{1(j)}')(X_{il}^* - \zeta_{1(l)}')' - \zeta_{2(j,l)}\right\}_{j,l}
    \end{bmatrix}\right).\]

    The resultant estimating equations for the parameters of interest then plug-in the corresponding $\zeta_{\cdot}$ for $\mu_j$, $\Sigma_{X_j^*X_l^*}$, or $\Sigma_{X_j^*X_j^*}$, and then are included as expressed above. This can be expressed as 
    \begin{equation} 
        g_i(\cdot) = \begin{bmatrix} \left\{X_{ij}^* - \xi_{1(j)}\right\}_{j} \\ 
            \left\{(X_{ij}^* - \xi_{1(j)})(X_{il}^* - \xi_{1(l)}) - \xi_{2(j,l)}\right\}_{(j,l)} \\ 
            |J_{0}|^{-1}\left[\sum_{j=1}^k\left(I(j\in J_{01}) + I(j\in J_{0}\setminus J_{1})\xi_{1(j)}\right)\xi_{6(j)}\right] - \xi_3 \\ 
            k^{-1}\sum_{j=1}^k\xi_{6(j)}^{-1}\xi_{8(j)} - \xi_4 \\
            \left\{\xi_{1(j)} - \xi_{6(j)}\xi_3 - \xi_{5(j)}\right\}_{j} \\ 
            \left\{(k-1)^{-1}\sum_{l\neq j; l=1}^k \xi_{2(j,l)}\xi_{8(l)}^{-1} - \xi_{6(j)}\right\}_{j} \\ 
            \left\{\xi_{2(j,j)} - \xi_{6(j)}^2\xi_3 - \xi_{7(j)}\right\} \\ 
            \left\{\sum_{l\neq j; l=1}^kI(l\in J_1)|J_1|^{-1}\xi_{2(l,j)} - \xi_{8(j)}\right\}_{j} \end{bmatrix}\label{eq::estimating_eq_form}
    \end{equation}
\end{proof}
\begin{proof}[Proof of Lemma \ref{lemma::general_asymptotic_distribution}]
    Taking the definitions as stated in the Lemma, note that we have the parameter vector $\widehat{\Theta}_* = (\widehat{\Theta}', \widehat{\xi}')'$ solves the equation given by \[0 = n^{-1}\sum_{i=1}^n\begin{bmatrix}
        U_n(\widehat{\Theta}, \widehat{\xi}) \\
        g_n(\widehat{\xi})
    \end{bmatrix}, \] and as a result we have \[\sqrt{n}\left(\widehat{\Theta}_* - \Theta_*\right) \stackrel{d}{\to} N\left(\mathbf{0}, \A^{-1}(\Theta,\xi)\B(\Theta,\xi)\A^{-1}(\Theta,\xi)'\right). \] All that's left is then to note that $\sqrt{n}(\widehat{\Theta} - \Theta) = \sqrt{n}(Q\widehat{\Theta}_* - Q\Theta_*)$, and so a standard application of the Delta Method gives the necessary result. Note that the specific forms give \begin{align*}
        A_{1,1}(\Theta,\xi) &= \frac{\partial}{\partial\Theta'}U_n(\Theta,\xi) &&A_{1,2}(\Theta,\xi) = \frac{\partial}{\partial\xi'}U_n(\Theta,\xi) \\
        A_{2,1}(\Theta,\xi) &= \frac{\partial}{\partial\Theta'}g_n(\xi) = \mathbf{0} &&A_{2,2}(\Theta,\xi) = \frac{\partial}{\partial\xi'}g_n(\xi) \\
        B_{1,1}(\Theta,\xi) &= U_n(\Theta,\xi)U_n(\Theta,\xi)' &&B_{1,2}(\Theta,\xi) = U_n(\Theta,\xi)g_n(\xi)' \\
        B_{2,1}(\Theta,\xi) &= g_n(\xi)U_n(\Theta,\xi)' &&B_{2,2}(\Theta,\xi) = g_n(\xi)g_n(\xi)' \\
        \A(\Theta,\xi) &= E\left\{\begin{bmatrix} A_{1,1} & A_{1,2} \\ A_{2,1} & A_{2,2} \end{bmatrix}\right\} &&\B(\Theta,\xi) = E\left\{\begin{bmatrix} B_{1,1} & B_{1,2} \\ B_{2,1} & B_{2,2} \end{bmatrix}\right\}. \end{align*} 
\end{proof}
\begin{proof}[Proof for Lemma \ref{lemma::conditional_mean}]
    While (a) was demonstrated in the proof for Lemma A.1 \cite{Carroll1990}, we include the full detail here as it is instructive for proving (b). 

    First note that, when $\delta = 0$ we get $V_1 = V_2$ in both (a) and (b). As a result, for $\delta \approx 0$, an arbitrary function of $V_1$, $h_{V_1}(v)$ is such that $h_{V_1}(v) = h_{V_2}(v) + O_p(\delta)$, from a first-order Taylor expansion. This gives the results for both covariance terms. It also gives the fact that, in either scenario, we can write $\frac{f_{V_2}'(v)}{f_{V_2}(v)} = \frac{f_{V_1}'(v)}{f_{V_1}(v)} + O_p(\delta)$. Then,   \begin{align*}
        E\left[\left.V_3\right|V_1\right] &= \frac{1}{f_{V_1}(V_1)}\int v_3f_{V_1,V_3}(V_1,v_3)dv_3 \\
        &\begin{cases}
            \stackrel{\text{(a)}}{=} \frac{1}{f_{V_1}(V_1)}\int v_3f_{V_3|V_2}(v_3|V_2=V_1-\delta v_3)f_{V_2}(V_1 - \delta v_3)dv_3, \\
            \stackrel{\text{(b)}}{=} \frac{1}{f_{V_1}(V_1)}\int v_3(\mathbf{1} + \delta v_3)^{-1}f_{V_3|V_2}(v_3|V_2=V_1(\mathbf{1} + \delta v_3)^{-1})f_{V_2}(V_1(\mathbf{1} + \delta v_3)^{-1})dv_3.
        \end{cases}
    \end{align*} The remainder of the proof follows by considering Taylor expansions of the integrands around $\delta = 0$, and noting that $f_{V_1}(V_1) = f_{V_2}(V_1) + O_p(\delta)$. Taking first the expression for (a), note that evaluating the expression at $\delta = 0$ gives $v_3f_{V_3|V_2}(v_3|V_2=V_1)f_{V_2}(V_1)$, which integrating gives $E[V_3|V_2=V_1]f_{V_2}(V_1) = 0$ by assumption. Differentiating the integrand, and evaluating at $\delta = 0$ gives the expression \[-v_3v_3'f_{V_3|V_2}'(v_3|V_2=V_1)f_{V_2}(V_1) - v_3v_3'f_{V_3|V_2}(v_3|V_2=V_1)f'_{V_2}(V_1),\] where the prime on the conditional density represents the derivative with respect to the conditioning term. Integrating these terms gives \[-\left[\Tr\left\{\frac{\partial}{\partial v_1}\Omega(v_1)\right\}f_{V_2}(V_1) + \Omega(V_1)f_{V_2}'(v_1) \right]_{v_1=V_1}. \] Combining this with the Taylor expansion in the denominator gives the desired result. 

    For (b) we follow a similar strategy. The integral evaluates to $0$ when $\delta = 0$ (by assumption), and the first derivative of the integrand with $\delta = 0$ is given by \[-v_3\circ v_3f_{V_3|V_2}(V_3|V_2=V_1)f_{V_2}(V_1) - v_3v_3'f'_{V_3|V_2}(V_3|V_2=V_1) - v_3v_3'f_{V_3|V_2}(V_3|V_2=V_1)f_{V_2}'(V_1). \] Once again, we integrate giving \[-\left[\text{diag}\left\{\Omega(v_1)\right\}f_{V_2}(V_1) + \Tr\left\{\frac{\partial}{\partial v_1}\Omega(v_1)\right\}f_{V_2}(V_1) + \Omega(V_1)f_{V_2}'(v_1)\right]_{v_1=V_1}. \] Then expanding the denominator as with (a) gives us the necessary result.
\end{proof}
\begin{proof}[Proof of Theorem \ref{thm::conditional_means}]
    This theorem follows as a direct application of Lemma \ref{lemma::conditional_mean}. For the additive case, we consider $V_1 \equiv X^*$, $V_2 \equiv \eta_0 + \eta_1X$, and $V_3 \equiv U$. Then, it is clear that $E[U|\eta_0 + \eta_1 X] = 0$, by our outlined assumptions, and as a result, $E[U|X^*] = -\delta\left[\Tr\left(\frac{\partial\Omega(x)}{\partial x}\right) + \Omega(x)\frac{f_{X^*}'(x)}{f_{X^*}(x)}\right]_{x=X^*} + O_p(\delta^2)$. Now, since $X = \eta_1^{-1}\left(X^* - \eta_0 - \delta U\right)$, the results follows directly. The multiplicative case requires additional considerations, but is otherwise similar. 

    First, taking $V_1 \equiv X^* - \eta_0$, $V_2 \equiv \eta_1X$, and $V_3 \equiv U$, then we note that $E[U|X^*=x] = E[V_3|V_1=x-\eta_0]$ and $\cov(U|X^*=x) = \cov(V_3|V_1=x-\eta_0)$. Additionally, $f_{V_1}(v) = f_{X^*}(v+\eta_0)$. Then, in order to solve for $E[X|X^*]$, we make use of a Taylor expansion of $X = (\mathbf{1} + \delta U)^{-1}(X^* - \eta_0)$, around $\delta = 0$, to handle the ratio. In particular, we consider the second order expansion so as to maintain an error of order $O_p(\delta^3)$ overall. That is, consider \begin{align*}
        (1 + \delta U)^{-1} &= 1 - \delta U + \delta^2\text{diag}\left(UU'\right) + O_p(\delta^3) \\
        \implies &E\left[\left.(1 + \delta U)^{-1}\right|X^*\right] \\
        &= 1 - \delta E[U|X^*] + \delta^2\text{diag}\left(\cov(U|X^*) + E[U|X^*]E[U|X^*]'\right) + O_p(\delta^3) \\
        &= 1 - \delta E[U|X^*] + \delta^2\text{diag}\left(\cov(U|X^*)\right) + O_p(\delta^3),
    \end{align*} where the last equality holds since $\delta^2E[U|X^*]E[U|X^*]' = O_p(\delta^4)$. Then, noting that \begin{align*}
        &E[U|X^*=x] \\
        &= E[V_3|V_1=x-\eta_0]\\ 
        &= -\delta\left[\text{\normalfont diag}\left\{\Omega(v_1)\right\} + v_1\circ\Tr\left\{\frac{\partial}{\partial v_1}\Omega(v_1)\right\} + v_1\circ\Omega(v_1)\frac{f_{V_1}'(v_1)}{f_{V_1}(v_1)}\right]_{v_1=x-\eta_0} + O_p(\delta^2) \\
        &= -\delta\left[\text{\normalfont diag}\left\{\Omega(x - \eta_0)\right\} + (x - \eta_0)\circ\left\{\left.\Tr\left(\frac{\partial\Omega(v)}{\partial v}\right)\right|_{v=x-\eta_0} + \Omega(x-\eta_0)\frac{f_{X^*}'(x)}{f_{X^*}(x)}\right\}\right] \\
        &+ O_p(\delta^2).
    \end{align*} Combining these two quantities gives the desired result.
\end{proof}
\begin{proof}[Proof of Theorem \ref{thm::asymptotic_normality_rc}]
    The proof will be presented, where convenient, using notation that implies scalar $X$. This can be extended to the multivariate case by carefully vectorizing the relevant M-estimators. First, note that $\widehat{\Theta}_\text{RC}$ solves $U_n(Y,Z,\widehat{X},\widehat{\Theta}_\text{RC}) = 0$. Now, $\widehat{X} = \widehat{\mu} + \widehat{\beta}\sum_{j=1}^k\widehat{\alpha}_kX^* + \widehat{\gamma}Z$ where the estimators $(\widehat{\mu}, \widehat{\beta}, \widehat{\gamma}, \{\widehat{\alpha}_j\}_j)$ solve \[h\left(\widehat{\mu},\widehat{\beta},\widehat{\gamma}, \{\widehat{\alpha}_j\}_j\right) = \begin{pmatrix}
        \mu_X - \widehat{\mu} - \widehat{\beta}\mu_{X^*} - \widehat{\gamma}\mu_Z \\
        \Sigma_{XX^*} - \widehat{\mu}\mu_{X^*} - \widehat{\beta}\Sigma_{X^*X^*} - \widehat{\gamma}\Sigma_{ZX^*} \\
        \Sigma_{XZ} - \widehat{\mu}\mu_{Z} - \widehat{\beta}\Sigma_{X^*Z} - \widehat{\gamma}\Sigma_{ZZ} \\
        \Tr\left\{\Sigma_{XX_j^*} - \widehat{\beta}\Sigma_{X^*X_j^*} - \widehat{\gamma}\Sigma_{ZX_j^*}\right\}_{j=1}^k
    \end{pmatrix} = \begin{pmatrix}
        0\\0\\0 \\ \{0\}_{j=1}^k
    \end{pmatrix}.\] Here, the reliance of the first three components on $\alpha_j$ is suppressed in $X^*$. By Lemma \ref{lemma::m_estimator_parameters} each of these components are estimable using an M-estimator. The previous results frame $\mu_{j}$, $\Sigma_{X_j^*X_l^*}$, $\Sigma_{XX_j^*}$, and $\Sigma_{X_j^*Z}$ in place of $\mu_{X^*}$, $\Sigma_{X^*X^*}$, $\Sigma_{XX^*}$, $\Sigma_{ZX^*}$, and $\Sigma_{X^*X_j^*}$. However, the latter can be written as transformations of the former. In particular, \begin{align*}
        \mu_{X^*} &= \sum_{j=1}^k \alpha_j\mu_j &&\Sigma_{X^*X^*} = \sum_{j=1}^k\sum_{l=1}^k \alpha_j\alpha_l\Sigma_{X_j^*X_l^*} &&&\Sigma_{XX^*} = \sum_{j=1}^k\alpha_j\Sigma_{XX_j^*} \\
        \Sigma_{X^*Z} &= \sum_{j=1}^k\alpha_j\Sigma_{X_j^*Z} &&\Sigma_{X^*X_j^*} = \sum_{l=1}^k\alpha_l\Sigma_{X_l^*X_j^*}.
    \end{align*} Now, noting that $h\left(\widehat{\mu},\widehat{\beta},\widehat{\gamma}, \widehat{\alpha}\right) = \mathbf{0} \iff n^{-1}\sum_{i=1}^n h\left(\widehat{\mu},\widehat{\beta},\widehat{\gamma}, \widehat{\alpha}\right) = \mathbf{0}$, this means that we can stack $g_i(\cdot)$ with $h(\cdot)$ which forms an estimating equation for the relevant parameters. Then, this can be stacked with $\Psi(\cdot)$, as the estimator of $\Theta_\text{RC}$ is given as solution to $n^{-1}\sum_{i=1}^n \Psi(Y_i,Z_i,\widehat{\mu}+\widehat{\beta}X_i^*+\widehat{\gamma}Z_i,\widehat{\Theta}_\text{RC}) = 0$. As a result, the asymptotic distribution of $\widehat{\Theta}_\text{RC}$ can be derived through the standard theory, using the M-estimator \[n^{-1}\sum_{i=1}^n \begin{pmatrix}
        \Psi(Y_i,Z_i,\widehat{\mu}+\widehat{\beta}X_i^*(\widehat{\alpha})+\widehat{\gamma}Z_i,\widehat{\Theta}_\text{RC}) \\
        h(\widehat{\mu},\widehat{\beta},\widehat{\gamma},\widehat{\xi}, \widehat{\alpha}) \\
        g_i(\widehat{\xi}) 
    \end{pmatrix} = \mathbf{0}.\] For the first matrix in the asymptotic covariance, denoted $\A_\text{RC}$, defined as the expectation of the $3\times3$ block matrix given by the derivatives of the previous estimating equation. In particular, we note that this will be an upper triangular matrix since $h$ is independent of $\Theta$, and $g_i$ is independent of $\Theta$ and $(\mu,\beta,\gamma, \alpha)$. Of course the precise form of this matrix will rely on the estimating equation for $\Theta$, and on the data available defining $g_i$. Generally, \begin{align*}
        \A_\text{RC} &= \begin{bmatrix}
        \A_{RC}^{(1,1)} & \A_{RC}^{(1,2)} & \A_{RC}^{(1,3)} \\
        \mathbf{0} & \A_{RC}^{(2,3)} & \A_{RC}^{(2,3)} \\
        \mathbf{0} & \mathbf{0} & \A_{RC}^{(3,3)}
    \end{bmatrix} \\
    &= \begin{bmatrix}
        E\left[\Psi_{\Theta}(\Theta,\mu,\beta,\gamma,\alpha,\xi)\right] & E\left[\Psi_{(\mu,\beta,\gamma,\alpha)}(\Theta,\mu,\beta,\gamma,\alpha,\xi)\right] & E\left[\Psi_{\xi}(\Theta,\mu,\beta,\gamma,\alpha,\xi)\right] \\
        \mathbf{0} & E\left[h_{(\mu,\beta,\gamma,\alpha)}(\mu,\beta,\gamma,\alpha,\xi)\right] & E\left[h_{\xi}(\mu,\beta,\gamma,\alpha,\xi)\right] \\
        \mathbf{0} & \mathbf{0} & E\left[g_{\xi}(\xi)\right]
    \end{bmatrix}, \end{align*} with $W_{\Delta}(\Delta)$ representing the derivative of $W(\cdot)$ with respect to $\Delta'$. Similarly,\[\B_\text{RC} = \begin{bmatrix}
        B_\text{RC}^{(1,1)} & B_\text{RC}^{(1,2)} & B_\text{RC}^{(1,3)} \\
        B_\text{RC}^{(2,1)} & B_\text{RC}^{(2,2)} & B_\text{RC}^{(1,3)} \\
        B_\text{RC}^{(3,1)} & B_\text{RC}^{(3,2)} & B_\text{RC}^{(3,3)}
    \end{bmatrix} = \begin{bmatrix}
        B_\text{RC}^{(1,1)} & B_\text{RC}^{(1,2)} & B_\text{RC}^{(1,3)} \\
        B_\text{RC}^{(2,1)} & B_\text{RC}^{(2,2)} & B_\text{RC}^{(1,3)} \\
        B_\text{RC}^{(3,1)} & B_\text{RC}^{(3,2)} & B_\text{RC}^{(3,3)}
    \end{bmatrix} = \begin{bmatrix}
        E[\Psi\Psi'] & \mathbf{0} & E[\Psi g'] \\
        \mathbf{0} & E[hh'] & \mathbf{0} \\
        E[g\Psi'] & \mathbf{0} & E[gg']
    \end{bmatrix},\] where the zeros come from noting that, since $E[\Psi] = E[g] = \mathbf{0}$, and that $h$ is constant (with respect to the underlying random variables), we have that $E[\Psi h'] = E[gh'] = \mathbf{0}$. Note that, in fact, the structure of $g$ is such that many of the components in the top right (and by symmetry bottom left) will also have this zero property, though, upon specification of $g$ this should become obvious. The standard theory of M-estimators then gives the asymptotic covariance of the stacked estimator as $\A_\text{RC}^{-1}\B_\text{RC}\A_\text{RC}^{-'}$. 
\end{proof}
\begin{proof}[Proof of Theorem \ref{thm::simex_asymptotic_normality}]
    The two proposed estimators for the SIMEX correction -- whether averaged before or after extrapolation -- can have their asymptotic distribution derived as an extension of \cite{SIMEX_asym} and Lemma \ref{lemma::general_asymptotic_distribution}. Our primary interest lies in $\widehat{\Theta} = \G(-1, \widehat{\Gamma})$, where $\widehat{\Gamma}$ is the parameter vector that minimizes $R(\Gamma)'C^{-1}R(\Gamma)$. Here, $C$ is a positive-definite matrix, decided on by the analyst (for instance, $C = I$ for standard least squares), $R(\Gamma) = \widehat{\Theta}_\Lambda - \G(\Lambda, \Gamma)$, and $\widehat{\Theta}_\Lambda$ is the vector formed by stacking $(\widehat{\Theta}_{\lambda_1}, \hdots, \widehat{\Theta}_{\lambda_R})$. $\widehat{\Theta}_\lambda$ is given by $B^{-1}\sum_{b=1}^B \widehat{\Theta}_{b,\lambda}$ for each $\lambda \in \Lambda$, and $\widehat{\Theta}_{b,\lambda}$ solves $n^{-1}\sum_{i=1}^n \psi(Y_i,Z_i,X_{bi}^*(\lambda),\Theta_\lambda) = 0$. Thus, we work to derive the asymptotic distribution of $\sqrt{n}(\widehat{\Gamma}-\Gamma)$, and then apply the Delta method for the necessary results.
    
    Note that, by definition we have $\Theta_\Lambda = \G(\Gamma,\Lambda)$ and $\widehat{\Theta}_\Lambda = \G(\widehat{\Gamma}, \Lambda)$. Define $s(\Gamma) = \frac{\partial}{\partial\Gamma}\G(\Gamma,\Lambda)'$. A Taylor expansion of $\G$ results in $\G(\widehat{\Gamma},\Lambda) = \G(\Gamma,\Lambda) + s(\Gamma)'\left\{\widehat{\Gamma} - \Gamma\right\} + o_p(1)$, which re-arranging and multiplying by $\sqrt{n}s(\Gamma)C^{-1}$ (for invertibility), and defining $\Omega(\Gamma) = s(\Gamma)C^{-1}s(\Gamma)'$ gives that \[\sqrt{n}\left(\widehat{\Gamma} - \Gamma\right) = \Omega(\Gamma)^{-1}s(\Gamma)C^{-1}\cdot\sqrt{n}\left(\widehat{\Theta}_\Lambda - \Theta_\Lambda\right) + o_p(1). \] As a result, we can focus the proof on the asymptotic distribution of $\sqrt{n}\left(\widehat{\Theta}_\Lambda - \Theta_\Lambda\right)$, and then apply a straightforward transformation for the distribution of $\widehat{\Gamma}$.
    
    For the estimator computed as the average after extrapolation, we focus on $\widehat{\Theta}_\text{SIMEX}^{(j)}(\lambda)$, which use Equation \ref{eq::modified_simex_form} directly for error term $j$. Stacking each of these estimators over the values of $\lambda\in\Lambda$, we get $\widehat{\Theta}_\Lambda^{(j)}$, and then consider the stacked version, stacking over $j=1,\hdots,k$, to be given by $\widehat{\Theta}_\Lambda$. This notation must be extended to the other relevant parameters: $\Gamma_{j}$ for the $j$-th extrapolant values, giving $\Omega(\Gamma_j) = s_j(\Gamma_j)C_j^{-1}s_j(\Gamma_j)'$. Then, the transformations here apply for each $j$, and so $\Omega(\Gamma)$, $s(\Gamma)$, and $C$ are formed by taking the block diagonal matrices over all $j$. With these amendments, the following argument applies directly. Once joint estimators are obtained for each $\G(-1, \Gamma_j)$, the final distribution can be taken by applying the relevant averaging transformation.

    An asymptotic linearization of $n^{-1}\sum_{i=1}^n\psi(Y_i,Z_i,X_{bi}^*(\lambda),\Theta_\lambda) = 0$ leads to \[\sqrt{n}\left(\widehat{\Theta}_{b,\lambda} - \Theta_\lambda\right) = \A^{(1,1)^{-1}}(\lambda)\sqrt{n}\sum_{i=1}^n\psi(Y_i,Z_i,X_{bi}^*(\lambda),\Theta_\lambda) + o_p(1),\] where $\A^{(1,1)^{-1}}(\lambda) = E\left[\frac{\partial}{\partial\Theta'}\psi(Y,Z,X_{b}^*(\lambda),\Theta_\lambda)\right]$. Then, averaging both sides over $b$, results in $\sqrt{n}\left(\widehat{\Theta}_{\lambda} - \Theta_\lambda\right) = \A^{(1,1)^{-1}}(\lambda)\sqrt{n}\sum_{i=1}^nB^{-1}\sum_{b=1}^B\psi(Y_i,Z_i,X_{bi}^*(\lambda),\Theta_\lambda) + o_p(1)$. This result holds for all $\lambda \in \Lambda$, where $\Lambda$ is taken to be the fixed grid of size $R$ that we simulate at. 
    
    The computation of these estimators, however, rely on the components of $\xi$ identified in Lemma \ref{lemma::m_estimator_parameters}, through the estimating equation $n^{-1}\sum_{i=1}^ng_i(\cdot) = 0$, and on the weights $\alpha$. We specify an M-estimator for each $\alpha_j$, based on some optimality criteria, and include the weights $\alpha_j$ in $\xi$. For both estimators under consideration, all parameters required for correction are then contained in $\xi$, and we can write \[n^{-1}\sum_{i=1}^n\psi(Y_i,Z_i,X_{bi}^*(\lambda),\Theta_\lambda) = n^{-1}\sum_{i=1}^n\psi\left(Y_i,Z_i,\eta_{1\cdot}^{-1}\circ\left[X_{i}^*-\eta_{0\cdot}+\sqrt{\lambda}M_*^{\frac{1}{2}}\nu_b\right],\Theta_\lambda\right), \] which we define to be $n^{-1}\sum_{i=1}^n\psi_{ib}(\lambda)$, where the necessary alterations are made to have this stacked over $j$ as discussed above. Writing the joint M-estimator as $n^{-1}\sum_{i=1}^n \begin{bmatrix}\psi_{ib}(\lambda) \\g_i \end{bmatrix} = 0$, and applying the exact argument as above, we get that \begin{align*}\sqrt{n}\left(\begin{bmatrix}
        \widehat{\Theta}_\lambda \\ \widehat{\xi}
    \end{bmatrix} - \begin{bmatrix}
        \Theta_\lambda \\ \xi 
    \end{bmatrix}\right) &= \A^{-1}(\lambda)\sqrt{n}\sum_{i=1}^nB^{-1}\sum_{b=1}^B\begin{bmatrix}
        \psi_{ib}(\lambda) \\ g_i
    \end{bmatrix} + o_p(1) \\
    &\triangleq \A^{-1}(\lambda)\sqrt{n}\sum_{i=1}^n\begin{bmatrix}
        \Psi_i(\lambda) \\ g_i
    \end{bmatrix} + o_p(1), \end{align*} where, again, the last equality is taken to be a notational definition. We have that \[\A(\lambda) = \begin{bmatrix}
        A^{(1,1)} & A^{(1,2)} \\
        \mathbf{0} & A^{(2,2)}
    \end{bmatrix} = \begin{bmatrix}
        E\left[\frac{\partial}{\partial\Theta'}\psi(Y,Z,X_b^*(\lambda),\Theta_\lambda)\right] & E\left[\frac{\partial}{\partial\xi'}\psi(Y,Z,X_b^*(\lambda),\Theta_\lambda)\right] \\
        \mathbf{0} & E\left[\frac{\partial}{\partial\xi'}g(\xi)\right]
    \end{bmatrix}.\]
    
    Define $\Theta_\Lambda$ to be the vector stacking $(\Theta_{\lambda_1}, \Theta_{\lambda_2},\hdots,\Theta_{\lambda_R})$, with the corresponding definition for $\widehat{\Theta}_{\Lambda}$, $\widetilde{\Psi}_{i}(\Lambda)$ to be the vector stacking $(\Psi_{i}(\lambda_1),\hdots,\Psi_{i}(\lambda_R),g_i)$, and $\A_\text{SIMEX}({\Lambda})$ to be the matrix with $\A^{(1,1)}(\lambda_1), \A^{(1,1)}(\lambda_2), \hdots, \A^{(1,1)}(\lambda_R)$ on the diagonals first $R$ diagonals, and then an $R+1$ column with $(\A^{(1,2)}(\lambda_1), \hdots, \A^{(1,2)}(\lambda_R), \A^{(2,2)})$, then zeros elsewhere. Note that the $\A^{(2,2)}$ portion of the matrix is constant across all $\lambda$, and so this matrix forms a block upper triangular matrix, with $(R+1)\times(R+1)$ blocks; each row $j$ takes the relevant matrix from $\A(\lambda_j)$ in the $j$-th block, and takes the cross matrix in the $R+1$ block. The above result implies that \[\sqrt{n}\left(\begin{bmatrix}
        \widehat{\Theta}_\Lambda \\
        \widehat{\xi}
    \end{bmatrix} - \begin{bmatrix}
        \Theta_\Lambda \\
        \xi
    \end{bmatrix}\right) = \A^{-1}(\lambda)\sqrt{n}\sum_{i=1}^n\widetilde{\Psi}_i(\Lambda) + o_p(1).\] Standard asymptotic theory then gives that this converges in distribution to a mean zero normal distribution, with variance given by $\A^{-1}(\Lambda)\Sigma\A^{-1}(\Lambda)'$, where $\Sigma = E\left[\widetilde{\Psi}(\Lambda)\widetilde{\Psi}(\Lambda)'\right]$. To extract only the distribution of $\sqrt{n}\left(\widehat{\Theta}_\Lambda - \Theta_\Lambda\right)$, we multiply by \[Q = \begin{bmatrix}I_{\dim\Theta_\Lambda\times\dim\Theta_\Lambda} & \mathbf{0}_{\dim\Theta_\lambda\times\dim\xi}\end{bmatrix}\], giving the same mean zero with covariance $Q\A^{-1}(\Lambda)\Sigma\A^{-1}(\Lambda)'Q'$. Combining this with the previous discussion gives \[\sqrt{n}\left(\widehat{\Gamma} - \Gamma\right) \stackrel{d}{\to} N\left(0,\Sigma_*\right), \] where $\Sigma_* = \Omega^{-1}(\Gamma)s(\Gamma)C^{-1}Q\A^{-1}(\Lambda)\Sigma\A^{-1}(\Lambda)'Q'C^{-1'}s(\Gamma)'\Omega^{-1}(\Gamma)'$.

    Finally, the SIMEX estimators are defined by taking the estimated $\widehat{\Gamma}$, and plugging into $\G$ at $\lambda = -1$. As a result, we apply the Delta method with $\G(-1, \cdot)$ as the function, and (assuming that it satisfies the requisite properties) we find that \[\sqrt{n}\left(\G\left(\widehat{\Gamma}, -1\right) - \G\left(\Gamma, -1\right)\right) \stackrel{d}{\to} N\left(\mathbf{0}, \G_{\Gamma}(-1,\Gamma)\Sigma_*\G_{\Gamma}(-1,\Gamma)'\right).\]

    If using the estimator which has been averaged prior to extrapolation, this gives us the required distribution. Otherwise, this has resulted in a $(k\dim\Theta)\times 1$ stacked estimator $\G(-1,\widehat{\Gamma})$, and so $\widehat{\Theta}_\text{SIMEX}$ is given by multiplying through the matrix $Q^*$ which is given by $\begin{bmatrix} \alpha_1I_{\dim\Theta} & \cdots & \alpha_kI_{\dim\Theta}\end{bmatrix}$, where $\sum_{j=1}^k\alpha_j = 1$. This results in a final asymptotic covariance of $Q^*\G_{\Gamma}(-1,\Gamma)\Sigma_*\G_\Gamma(-1,\Gamma)'Q^{*'}$.

    While the notational conventions were the same for either the averaging before, or the averaging afterwards, we note that the matrix structures are fundamentally different between the two. This is true even before the adjustment with $Q^*$, since $\Omega(\Gamma)$, $s(\Gamma)$, $C$, $Q$, $\A$ and $\Sigma$ are all of different forms and shapes. 
\end{proof}

\section{Supplementary Appendix B}
In the moment reconstruction setup, the correction parameters need to be estimated using M-estimators. By assumption, $\Sigma_{XX}$ is both the conditional and unconditional variance of $X_i$, which means it is estimated in $\xi$. Further, $\eta_{l\cdot}$ contain only $\{\alpha_j\}$ and parameters estimated in $\xi$. This leaves $\cov(X_i^*|Y_i)$ and $E[X_i^*|Y_i]$ to be estimated. 

To do so, we can form standard joint M-estimators. For $\Theta_1 \equiv E[X_i^*|Y_i=1]$ we take $0 = n^{-1}\sum_{i=1}^ny_i\sum_{j=1}^K\alpha_jX_{ij}^* - \Theta_1$, $\Theta_2 \equiv E[X_i^*|Y_i=0]$ we take $0 = n^{-1}\sum_{i=1}^n(1-y_i)\sum_{j=1}^K\alpha_jX_{ij}^* - \Theta_2$, $\Theta_3 \equiv \cov(X_i^*|Y_i=1)$ we take $0 = n^{-1}\sum_{i=1}^ny_i\left(\sum_{j=1}^k\alpha_jX_{ij}^* - \Theta_1\right)^2 - \Theta_3$, and $\Theta_4 \equiv \cov(X_i^*|Y_i=0)$ we take $0 = n^{-1}\sum_{i=1}^n(1-y_i)\left(\sum_{j=1}^k\alpha_jX_{ij}^* - \Theta_2\right)^2 - \Theta_4$. We have assumed the $\{\alpha_j\}$ are fixed, however, they can be estimated as well, stacking in a similar way. 

\bibliographystyle{plainnat}
\bibliography{references}


\end{document}